\documentclass[12pt]{article}

\usepackage{fullpage}
\usepackage[centertags]{amsmath}
\usepackage{amsfonts}
\usepackage{amssymb}
\usepackage{amsthm}
\usepackage{newlfont}
\usepackage{mathrsfs}
\usepackage{euscript}
\usepackage{siunitx}
\usepackage{physics}
\usepackage{bbold}
\usepackage[utf8]{inputenc}
\usepackage{graphicx}

\usepackage{color}
\usepackage{floatrow}
\usepackage{hyperref}
\usepackage{subcaption}
\usepackage{cite}
\usepackage{authblk}

\usepackage{tikz}
\usetikzlibrary{3d,math}
\pgfmathsetseed{10}

\theoremstyle{plain}

\theoremstyle{definition}

\theoremstyle{remark}

\numberwithin{equation}{section}

\newcommand{\bbC}{{\mathbb C}}

\newcommand{\opunit}{\text{1}\kern-0.22em\text{l}}

\DeclareMathAlphabet{\mathpzc}{OT1}{pzc}{m}{it}

\newcommand{\id}{\textrm{d}}

\begin{document}
\title{{\Large{\bf Diffraction and interference with\\ run-and-tumble particles}}}

\author[*]{Christian Maes}
\author[*]{Kasper Meerts}
\author[*,$\dagger$]{Ward Struyve}
\affil[*]{Instituut voor Theoretische Fysica, KU Leuven}
\affil[$\dagger$]{Centrum voor Logica en Filosofie van de Wetenschappen, KU 
Leuven}
\date{}

\maketitle

\begin{abstract}
	Run-and-tumble particles, frequently considered today for modeling bacterial locomotion, naturally appear outside a biological context as well. 
Here, we consider them in a quantum mechanical relation, using a wave function to drive their propulsion and tumbling.  Such 
quantum-active motion realizes a jittery motion of Dirac electrons (as in the famous {\it 
Zitterbewegung}): the Dirac electron {\it is} a 
run-and-tumble particle, where the tumbling is between chiralities.  We visualize the electron trajectories in single and double slit 
experiments and discuss their dependence on the spin-direction. In particular, that yields the time-of-arrival statistics 
of the electrons at the screen. Finally, we observe that away from pure quantum 
guidance, run-and-tumble particles with suitable spacetime-dependent parameters
produce an interference pattern as well.
 \end{abstract}

\section{Introduction}
Active media make an important topic of current research efforts in nonequilibrium statistical mechanics. The mechanism of self-propulsion, e.g.\ by consuming stored or ambient energy, is central and combines interestingly with repulsive interactions showing collective behavior such as phase separation or flocking, \cite{1}.  For biophysical applications, models often imitate bacterial motion and biological motors.  Major examples include active Brownian motion \cite{ab1,ab2} and run-and-tumble models \cite{rtp}.  The latter describe particles (like {\it E.\ coli} in polymeric fluids \cite{pat})  with flagellar self-propulsion where the `free run' is interrupted by tumbling events where the orientation is updated.\\
 Such models of active media often have a longer history however; see e.g.\ \cite{wei}. The origin of run-and-tumble processes goes  back to the problem of electromagnetic wave propagation in a double transmission line (the transatlantic cable of the 1850's) such as summarized in the telegraph equation of Oliver Heaviside and William Thomson,  and to the corresponding telegraph process as pioneered by Mark Kac \cite{kac}. Choosing suitable boundary conditions \cite{masoliver}, independent-particle models simulating running and standing waves are easily obtained. In the present paper, we explore and visualize a newer relation with the Dirac theory  \cite{dir} of the electron. \\

According to the usual formulation of the Dirac theory, an electron is described by a spinor. By decomposing the Dirac spinor into its left- and right-handed chiral component, the theory can be seen as one of interacting Weyl particles. That is, the electron is viewed as a massless Weyl particle with a chirality that flips between left- and right-handed. That interpretation is strongly suggested in particular  by the Standard Model of particles physics, according to which electrons are fundamentally massless and hence Weyl particles, acquiring an effective mass through the interaction with the Higgs field. Roger Penrose describes this view in detail in \cite{pen} referring to it as the {\em zig-zag picture}, writing: ``So are these zigs and zags real?  For my own part, I would say so; they are as real as the ``Dirac electron'' is itself real...'' \cite[p.\ 632]{pen}. That perspective has in fact a longer history, probably starting with Richard Feynman who suggested and formulated a path-integral version for the Dirac equation in two-dimensional spacetime where the electron paths go back and forth at the speed of light \cite{feyn,schweb,cart}.\\ 

That zig-zagging of electrons, as they flip between right-handed and left-handed chirality, is made manifest in the pilot-wave theory and makes the topic of the present study. 
Then, most interestingly, the electron dynamics becomes that of a run-and-tumble particle with propulsion speed and tumbling rates depending on the left- and right-handed chirality component of the Dirac spinor. In that way, we connect an interesting model of nonequilibrium statistical mechanics, the active motion of run-and-tumble particles, with Dirac theory. As a result, we can visualize the electron dynamics in the case of diffraction and interference phenomena, studying the effect of the spin direction. The explicit connection with the (stochastic) run-and-tumble model indeed allows simulating the electron path in the nonrelativistic Pauli approximation. The availability of particle trajectories also allows a straightforward determination of arrival times on the screen after exiting the double-slit, something that is difficult to achieve otherwise \cite{muga2000}.
The numerical method employs a Cash-Karp algorithm in Julia \cite{cashkarp,julia}; see \cite{zigzag} for the publicly available code. \\



The outline of the paper is as follows. In the next section, we start with the general setup of run-and-tumble particles with spacetime-dependent propulsion speeds and tumbling rates, as a slight extension of the usual practise for active particles in nonequilibrium statistical mechanics. In Section \ref{zig}, we show how the run-and-tumble dynamics for a Dirac particle can be obtained in the pilot-wave approach. Following \cite{deA,cowi,ward}, this is done by identifying the continuity equation for the Dirac electron, when viewed as combination of chiral Weyl electrons, with the Master equation for run-and-tumble particles. The dynamics contains what we call a ``curling'' and a ``tumbling'' component, which both depend on the spin. We study the effects of these components, comparing 
with previous studies in \cite{philippidis1,philippidis2,gondran05,sanz08,sanz14,sanz15} where either the curling or the tumbling aspects were ignored. The main part of the present paper is the application of these ideas to the simulation of diffraction and interference experiments. This is done within the non-relativistic limit of the Dirac theory, which is given by the Pauli theory. In Section \ref{vis}, we visualize the trajectories of the electrons producing the typical patterns upon hitting the screen. The simulation allows to speak here precisely about arrival times as well, as we illustrate in Section \ref{rem}.  We add in Section \ref{awa} that diffraction and interference patterns can be obtained away from the strict quantum context, within the dynamics of run-and-tumble particles more generally.  Here also, an implicit relation between pure quantum processes such as in spin-orbit coupling and (classical) run-and-tumble particles was formulated before.  References with such analogies and similar  inspiration include \cite{loewe, burada}.  The recent \cite{bierkens} gives a mathematical (spectral) analysis of the  zig-zag process within classical probability. We conclude in Section \ref{con}.\\

\section{Spacetime activity}\label{set}
We consider independent active particles in three spatial dimensions, that alternatingly follow one of two spacetime-dependent velocity fields $\mathbf v_{\pm}(\mathbf x,t)$,
\begin{equation}\label{gui}
\dot{\mathbf x}_t = \mathbf{v}_{\chi_{{}_{t}}} (\mathbf x_t,t)
\end{equation}
for position $\mathbf x_t$ at time $t$.  The variable $\chi_{{}_t} =\pm$ (we think of it as chirality, with right-handed particles having $\chi=+$, left-handed particles having $\chi=-$) follows an inhomogeneous Markov jump process with transition rate $a(\mathbf x, t)$ to jump from $- \longrightarrow +$ and rate $b(\mathbf x, t)$ to jump $+\longrightarrow -$ when $\mathbf  x_t= \mathbf x$.  Then, the resulting Master equation for the densities $\rho(\mathbf x,t;\chi)$ of left- and right-handed movers is
\begin{equation}\label{inr}
\frac{\partial \rho}{\partial t}(\mathbf x,t;\chi) + \,\mathbf \nabla\cdot\big( \mathbf v_\chi(\mathbf x,t) \rho(\mathbf x,t;\chi) \big) = r(\mathbf x,t;-\chi)\rho(\mathbf x,t;-\chi) - r(\mathbf x,t;\chi)\rho(\mathbf x,t;\chi)
\end{equation}
for the tumbling field $r(\mathbf x,t;\pm)$ with $r(\mathbf x,t;\chi=-) = a(\mathbf x,t)$ and $r(\mathbf x,t;\chi=+) = b(\mathbf x,t)$. 
Initial conditions should be added for the initial distribution of the particles (e.g.\ at time $t=0$ near $\mathbf x=0$ in what follows).  That extension of the usual run-and-tumble dynamics is strictly continued in Section \ref{awa}.  The relation with quantum mechanics appears next.\\

As to be explained in the following section, the wave function (spinor) that solves the Dirac or Pauli equation will determine $\mathbf v_{\pm}(\mathbf x, t)$ and $r(\mathbf x,t;\chi)$, such that \eqref{inr} is exactly the continuity equation for the corresponding quantum probability density. If the initial positions of the particles are thrown according to the Born rule, they will move following \eqref{gui} with some initial $\chi_0$, such that at all times they follow the quantum statistics.\\  
In that sense we may speak about quantum-active motion where the ``fuel'' is provided by a quantum evolution, yet without breaking time-reversal invariance (more on that below).

\section{Zig-zag theory of the Dirac electron}\label{zig}

The Dirac equation is the relativistic version of the Schr\"odinger equation 
for electrons. We are interested in trajectories that can be associated with a 
Dirac electron, as pioneered by David Bohm \cite{bohm53c}, and in the 
connection with run-and-tumble dynamics.  We follow \cite{ward} for the choice 
of decomposition of the wave function with the resulting continuity equations 
for left-handed and right-handed (so called Weyl) electrons.  The basic idea is 
that all fermionic particles are fundamentally chiral and massless, as suggested by the Standard Model. The 
electron mass $m$ couples the Weyl electrons which results in the tumbling of 
chiralities; the tumbling rate will be seen to be proportional to $mc^2/\hbar$.  
Between any two consecutive tumbling events the electrons move at the speed $c$ 
of light, but the tumbling results in an average velocity smaller than the 
light speed.
We now proceed with the mathematical details for that decomposition in four spacetime 
dimensions.\\

 A single electron in four spacetime dimensions has a wave function $\psi(\mathbf{x}, t) \in \bbC^4$ satisfying the (free) Dirac equation
 \begin{equation}\label{die}
(i\hbar\gamma^\mu \partial_\mu - mc)\psi =0
 \end{equation}
written in terms of the gamma-matrices 
$\gamma^\mu$. Assuming the Weyl (or chiral) basis for the gamma matrices, the 
wave function can be written as a bispinor, a pair of two component Weyl spinors 
$\psi = (\psi_-,\psi_+)^T$, describing respectively a left-handed and a 
right-handed spinor. The Dirac equation \eqref{die} can then be rewritten as two 
coupled equations of these spinors,
\begin{equation}\label{wey}
  i\hbar\,\sigma^\mu \partial_\mu \psi_+ = mc\psi_-,\qquad   
  i\hbar\,\overline\sigma^\mu \partial_\mu \psi_- = mc\psi_+
\end{equation}
where $\sigma^\mu = (\mathbb{1},\vb*{\sigma})$ and $\overline\sigma^\mu = 
(\mathbb{1},-\vb*{\sigma})$ for Pauli matrices $\vb*{\sigma}$.
Thereupon, the Dirac current
\[
j^\mu = \overline{\psi}\gamma^\mu\psi\quad \text{with }\,\; \overline{\psi} = 
\psi^\dagger \gamma^0,\;\;\text{satisfying }\;\; \partial_\mu j^\mu=0
\]
follows the decomposition $j^\mu = j^\mu_+ + j^\mu_-$ for
\[
j^\mu_+ = \psi_+^\dagger \sigma^\mu\psi_+,\qquad  j^\mu_- = \psi_-^\dagger 
\overline\sigma^\mu\psi_-
\]
They do not satisfy a continuity equation separately, but have a source \[
\partial_\mu j^\mu_{\pm} = \pm F, \quad \text{with }\;\; F= 
2\frac{mc}{\hbar}\Im\psi_+^\dagger\psi_-
 \]
That last equation can be written out explicitly in the way of \eqref{inr},
\begin{eqnarray}\label{rt}
\partial_t\rho_+ + {\mathbf\nabla}\cdot({\mathbf v}_+\rho_+) &=& a\rho_- - b\rho_+\nonumber\\ 
\partial_t\rho_- + {\mathbf\nabla}\cdot({\mathbf v}_-\rho_-) &=& b\rho_+ - a\rho_-
\end{eqnarray} for the densities $\rho_+ = j^0_+ = \psi^\dagger_+\psi_+$ and 
$\rho_- = j^0_- = \psi^\dagger_-\psi_-$,
where
\begin{eqnarray}\label{rte}
\vb{v}_+ = c\,\frac{\psi_+^\dagger \,\vb*{\sigma}\, 
\psi_+}{\psi^\dagger_+\psi_+}, &&  \vb{v}_- = -c\, \frac{\psi_-^\dagger 
\,\vb*{\sigma}\,\psi_-}{\psi^\dagger_-\psi_-}\nonumber\\
a  = 2\frac{mc^2}{\hbar}\,\frac{\left( \Im\psi_+^\dagger 
\psi_-\right)^+}{\psi^\dagger_-\psi_-},&& b= 2\frac{mc^2}{\hbar}\,\frac{\left( 
\Im\psi_-^\dagger \psi_+\right)^+}{\psi^\dagger_+\psi_+}
\end{eqnarray} (equations (22)--(23) in \cite{ward}.)  Here we used that $F = 
F^+ - (-F)^+$ for $F^+ = \max\{F,0\}$. \\

As in the theory of stochastic processes, the natural next step is to interpret 
the above equations \eqref{rt}--\eqref{rte} as the Master equation for an ensemble of 
run-and-tumble particles, following the setup of Section \ref{set}.  They are here interpreted as two different massless 
manifestations of the Dirac electron, ``zig'' and ``zag,'' i.e., as the
Weyl electrons \cite{pen}.\\
 We consider therefore the stochastic dynamics of the electron position $\mathbf x_t$ as in \eqref{gui}.
The speed $|\mathbf v_{\pm}| = c$ is invariably equal to the speed of light.  (That follows from the fact that $j^\mu_+j_{+\mu} = j^\mu_-j_{-\mu} = 0$.)  What changes discontinuously at every transition $\chi\rightarrow -\chi$ is the direction of the particle, as the velocity changes from $\mathbf{v}_\chi$ to $\mathbf{v}_{-\chi}$.  The tumbling (and change of effective velocity field) is a consequence of the coupling between left- and right-chirality and indeed occurs at a rate proportional to the mass.  Note in any case that the factor $2mc^2/\hbar$ in the expressions \eqref{rte} for $a$ and $b$ is known as the angular frequency of the {\it Zitterbewegung}\footnote{The Zitterbewegung \cite{breit,shr,greiner90} is actually due to the interference of positive and negative energy wave functions (for the trajectories see \cite{hol,holland93b}). As such it should be distinguished from the zig-zag motion considered here which is due to the fundamental massless nature of the particles.}, here realized as the tumbling of the particle. The resulting dynamics is a time-dependent Markov process $(\mathbf x_t,\chi_t)$ and that dynamics is such as to reproduce at every moment the correct densities $\rho_{\pm}$. \\
The dynamics is also time-reversal invariant, in contrast with models of 
run-and-tumble particles where there is dissipation (in consuming fuel), 
\cite{fod}.  Given a trajectory $\omega= (\mathbf{x}(t),\chi_t), t\in [-s,s]$, 
its time-reversal $\theta \omega$ is $(\theta \omega)_t = 
(\mathbf{x}(-t),\chi_{-t}), t\in [-s,s]$.  If $\omega$ is a possible trajectory 
from applying the rules \eqref{rte} with solution $\psi(t,\mathbf x)$, then the 
time-reversed trajectory $\theta\omega$ can be generated from applying those 
rules to the solution  $i\gamma^1\gamma^3 \psi^*(\mathbf{x},t)$.  That 
time-reversal operation $\psi(\mathbf{x},t) \to  i\gamma^1\gamma^3 
\psi^*(\mathbf{x},t)$ \cite{greiner90}, or $\psi_\pm(\mathbf{x},t) \to  
-\sigma_2 \psi^*_\pm(\mathbf{x},t)$, is anti-unitary, and indeed implies that 
${\mathbf v}_\pm(\mathbf{x},t) \to -{\mathbf v}_\pm(\mathbf{x},-t)$.   Probabilities for trajectories arise here from the initial condition where the positions are drawn according to the Born rule and from the random tumbling events.  Since the transformation takes $|\rho_\pm(\mathbf{x},s)|^2 \to |\rho_\pm(\mathbf{x},-s)|^2$ and also maps $a(\mathbf{x},t) \to b(\mathbf{x},-t)$,
\, 
$b(\mathbf{x},t) \to a(\mathbf{x},-t)$, there is statistical reversibility \cite{duerr05a}, 
i.e.\ formally, the probability densities for $\omega$ and at $\theta\omega$ are equal.\\

To model the slit experiment it is natural to use Gaussian wave functions. 
However, for the Dirac theory the Gaussian solution is not of closed form \cite{greiner90}. For 
numerical simplicity, we therefore consider the non-relativistic limit of the 
Dirac theory \eqref{wey}-\eqref{rte}, given by the Pauli theory, which after all is realistic for 
the interference experiments with electrons as in the pioneering works \cite{it,ja}. We 
further assume that the spin decouples from the translational degrees of 
freedom, so that the Pauli spinor is of the form  $\phi(t, \vb{x}) = \psi(t, 
\vb{x})\, \xi$, where now $\psi$ is a scalar function governed by the 
Schrödinger equation, and $\xi$ a constant spinor.  
We define the spin polarization vector 
$\vu{s} = \xi^\dagger \vb*{\sigma} \xi$, where $\vb*{\sigma}$ are still the Pauli 
matrices.\\ Then, the zig-zag model goes as follows \cite{ward}. The velocity field for 
\eqref{gui} turns out to be{\footnote{In the non-relativistic limit actually an extra term appears in the velocity field \cite{ward}. However, this term makes that the distribution $|\psi|^2$ is not exactly preserved by the zig-zag dynamics, but only up to relevant order. Here, we have chosen to drop this term in order to guarantee preservation of the distribution $|\psi|^2$. See \cite{ward} for further discussion.}}
\begin{equation}
\vb{v}_\chi = 
 \frac{\hbar}{m} \Im \frac{\grad \psi}{\psi} + \frac{\hbar}{m} \Re 
\frac{\grad \psi}{\psi} \cross \vu{s} + \chi\, c \,\vu{s}\label{vf}
\end{equation}
That velocity field thus consists of three terms written in the right-hand side of \eqref{vf}. The first corresponds simply to the velocity field associated with solutions of the free Schr\"odinger equation\footnote{For a plane wave with wave vector $\mathbf k$, that first term in \eqref{vf} would be $\hbar \mathbf{k}/m$.}.  The second term with the curl produces a whirling orthogonal to the spin; we will refer to it as the ``curling''. The third term corresponds to the ``tumbling'' (as the chirality $\chi$ flips). This is the dominant one in \eqref{vf} as  being a light speed component in the direction of the spin vector.   We will see that between tumbling events the particle indeed approximately moves in straight lines at speed $c$.  The jumps between chiralities $\chi\rightarrow -\chi$ happen at a rate
\begin{equation}\label{eqn:rate}
r(\vb{x},t, \chi) = \left[\chi \tau(t,\vb{x}) \right]^+ 
\end{equation} where 
\begin{equation}\label{ta}
\tau(t,\vb{x}) = - c \frac{\grad{\abs{\psi}^2}}{\abs{\psi}^2}\vdot 
\vu{s} = -2  c \Re \frac{\grad\psi}{\psi}\vdot \vu{s}
\end{equation}
The trajectories of the Dirac electron are thus well approximated (for 
non-relativistic speeds) by that running and tumbling as summarized in 
\eqref{vf}--\eqref{ta} to be substituted in \eqref{gui}. (The dynamics is still 
time-reversal invariant, with now $\psi({\mathbf x}, t) \to \psi^*({\mathbf x}, 
-t)$ and $\xi \to -\sigma_2 \xi^* $.)

The dynamics without the curling and tumbling term corresponds to the one originally suggested by de Broglie and Bohm to describe non-relativistic particles without spin \cite{bohm93,holland93b}. Relativistic considerations suggest the addition of the spin-dependent curling term \cite{bohm93}, whereas the standard model suggests also the addition of the tumbling term \cite{cowi,ward}.

\section{Trajectory visualization}\label{vis}
The issue of visualization ({\it Anschaulichkeit}) 
has been subject of major discussion among the pioneers of quantum mechanics.  The ultimate goal, often ``refuted\footnote{It was for example the motivation for the Heisenberg paper (1927) on the uncertainty relations.}'', is to derive and to show the trajectories of quantum particles, thus shedding light on the underlying causal connections leading to  the empirical findings.  In the present section we visualize diffraction and interference with Pauli electrons, following the theoretical framework above.  

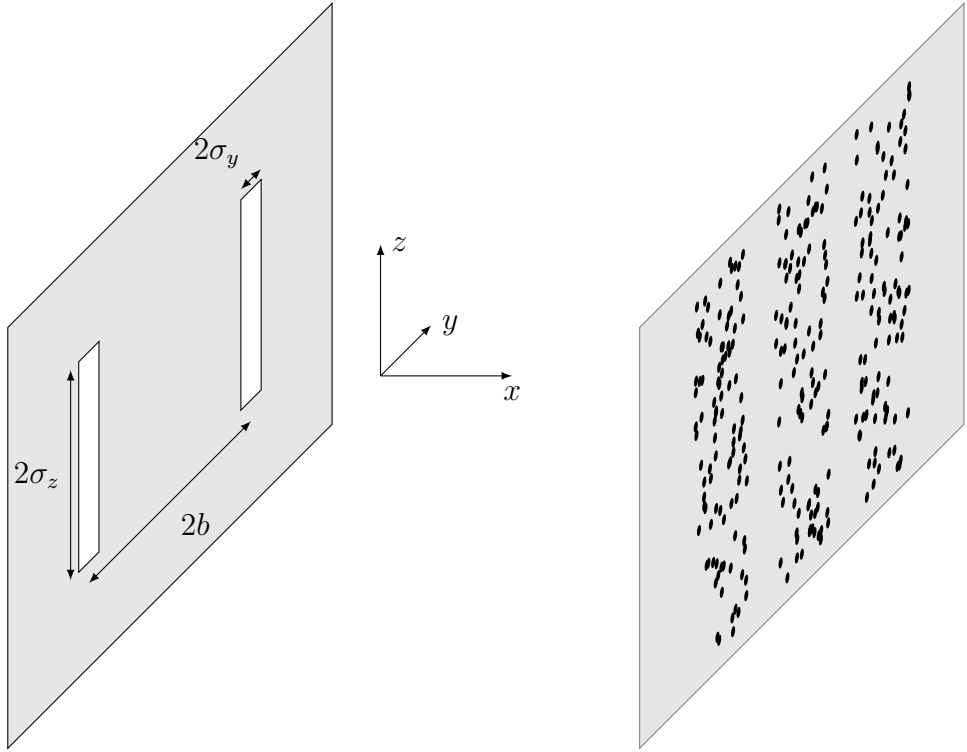
\begin{figure}
\begin{tikzpicture}[scale=0.70,>=latex]
\tikzmath{ \b = 4; \sy = 0.5; \sz = \sy * 4; \w = 2*\b; \h = 2*\sz; }
\begin{scope}[canvas is yz plane at x=0,rotate=-90]
	\draw[fill=black!10, even odd rule]
		(-\w,-\h) rectangle (+\w, +\h)
		(-\b-\sy,-\sz) rectangle (-\b+\sy,+\sz)
		(+\b-\sy,-\sz) rectangle (+\b+\sy,+\sz);
	\draw[<->] (+\b-\sy,1.1*\sz) -- (+\b+\sy,1.1*\sz) node[midway,above left] 
	{$2\sigma_y$};
	\draw[<->] (-\b-1.8*\sy,-1*\sz) -- (-\b-1.8*\sy,+1*\sz) node[midway,left] 
	{$2\sigma_z$};
	\draw[<->] (-\b,-1.2*\sz) -- (\b,-1.2*\sz) node[midway,below right] {$2b$};
\end{scope}
\begin{scope}[canvas is yz plane at x=12,rotate=-90]
	\draw[fill=black!20,opacity=0.5] (-\w,-\h) rectangle (+\w,+\h);
	\foreach \p in {1,...,100} {
		\fill ({\b * ( 0 + (rand)/3)},1.8*\sz*rand) circle (0.1);
		\fill ({\b * (+1 + (rand)/3)},1.8*\sz*rand) circle (0.1);
		\fill ({\b * (-1 + (rand)/3)},1.8*\sz*rand) circle (0.1);
	}
\end{scope}
\begin{scope}[xshift=4cm]
	\draw[->] (0,0,0) -- ++(2.5,0,0) node[below] {$x$};
	\draw[->] (0,0,0) -- ++(0,2.5,0) node[right] {$z$};
	\draw[->] (0,0,0) -- ++(0,0,-2.5) node[right] {$y$};
\end{scope}
\end{tikzpicture}
\caption{Spatial setup, with electrons exiting the wall from slits to arrive at the screen.}
\label{fig:setup}
\end{figure}

The physical setup for the simulated experiments is as seen in 
Fig.~\ref{fig:setup}. The Cartesian coordinates are such that the distance from 
the slit(s) to the screen is measured in the $x$-direction, the slits are 
separated from each other in the $y$-direction and the $z$-direction goes along the (vertical)
direction of the slit(s).\\

We model the wave function right after it exits the slit(s) as
\begin{equation}\label{ss}
\psi^G(t,\vb{x}\equiv(x,y,z)) = \psi_x(t,x) \psi^S(t,y) \psi_z(t,z),
\end{equation}
where 
\begin{equation}\label{ga}
\psi_x(t, x) = \frac{1}{\left(2\pi \sigma_x^2 \left(1 + i \frac{\hbar t}{2m 
		\sigma_x^2}\right)^2 \right)^{1/4}} \exp \left( -\frac{(x-v_0 t)^2}{4 \sigma_x^2 
	\left(1 + i \frac{\hbar t}{2m\sigma^2}\right)} + \frac{i}{\hbar} \left( m v_0 x 
- \frac{mv_0^2}{2} t \right) \right)
\end{equation}
with
\begin{equation}
\pdv{\psi_x}{x}(t,x) = \left(-\frac{x-v_0 t}{2\sigma_x^2 \left(1+i \frac{\hbar t}{2m\sigma_x^2}\right)} + \frac{i}{\hbar} m v_0 \right) \psi_x(t,x)
\end{equation}
is a Gaussian packet with standard deviation $\sigma_x$ moving with speed $v_0$ 
in the positive $x$-direction. We localize the wave function in the 
$x$-direction as much as possible, but not so much that the uncertainty in 
momentum space overwhelms the bulk velocity of the wave packet, i.e.\ $\Delta 
p_x \ll m v_0$.  The uncertainty principle $\Delta x \Delta p_x = \hbar/2$ 
implies therefore that $\sigma_x = \Delta x$ has a lower limit of a few times 
$\hbar / 2 m v_0$: we take $v_0=c/10$ and $\sigma_x = 50 \hbar / mc$.

For the $z$-direction, $\psi_z(t, z)$ is similarly a Gaussian wave packet with 
standard deviation $\sigma_z$, to be specified later, but with zero speed. The 
form of the packet $\psi^S(t,y)$ will depend on whether we consider one or two 
slits. 

The wave function $\psi^G$ is separable in the coordinates $x,y,z$. Therefore, if the spin is aligned along one of the coordinate axes, the motion in that direction will decouple from the motion in the other directions. For example, suppose $\vu{s} = \vu{x}$, then $ \vb{v}_\chi = (v_{\chi,x},v_{\chi,y},v_{\chi,z})$ equals
\begin{align}
v_{\chi,x} &=  \frac{\hbar}{m} \Im \left( \frac{1}{\psi_x} \pdv{\psi_x}{x} \right)   +  \chi\, c \label{xcomp}\\
v_{\chi,y} &=  \frac{\hbar}{m} \Im \left( \frac{1}{\psi^S} \pdv{\psi_y}{y} \right)   +   \frac{\hbar}{m} \Re \left( \frac{1}{\psi_z} \pdv{\psi_z}{z} \right)  \\
v_{\chi,z} &=  \frac{\hbar}{m} \Im \left( \frac{1}{\psi_z} \pdv{\psi_z}{z} \right)   -   \frac{\hbar}{m} \Re \left( \frac{1}{\psi^S} \pdv{\psi^S}{y} \right)  
\end{align}
The motion in the $y$-and $z$-direction are thus coupled and corresponds to the motion without tumbling. The motion in the $x$-direction has the tumbling but no curling. Hence, a plot of the projection of the trajectories in the $(y,z)$-plane would give the same trajectories with or without keeping the tumbling in the dynamics. When $\vu{s} = (\vu{x}+\vu{y}+\vu{z})/\sqrt{3}$, as we will often consider in the following, the motion along the different directions is coupled.

We are then ready to numerically integrate the stochastic dynamics described by 
\eqref{gui}.  We choose to implement the following algorithm in 
Julia, \cite{julia}. We employ the Cash-Karp method, \cite{cashkarp}, an adaptive 
Runge-Kutta method, giving a fourth order approximation with a fifth-order 
error estimate.  The latter is used to adaptively adjust the time step for the 
local truncation error to a reasonable value, say $10^{-12}$.  In addition, an 
upper limit is set for the product of the time step $\id t$ with the tumbling 
rate $r$: $r\,\id t < 2^{-7}$. If surpassed, the time step is lowered and the 
step is retried. For every Runge-Kutta step, the tumbling rate is sampled at 
the beginning and at the end of the step, the mean of these two values will 
then provide an approximation to the mean value of the rate over the step. If 
that mean tumbling rate does not exceed the threshold, the step is accepted, 
and a random number is chosen uniformly from 0 to 1. Only if the product of the 
time step and the rate exceeds that random number, the chirality of the particle 
is inverted. In either case, we continue on to the next step, until the 
particle hits the screen or the maximum number of iterations is reached.\\
The code is publicly available; see \cite{zigzag}.  The plots are made in 
natural units where $\hbar=c=m=1$. Since the distance traveled by the particle 
is usually much greater than the width of the wave function in the 
$y$-direction, we rescale the $y$-axis to comfortably fit all the data on the 
page. The same scale is used when different dynamics or spin orientations are 
explicitly compared.

\subsection{Single slit diffraction}
\begin{figure}
\centering
	\includegraphics{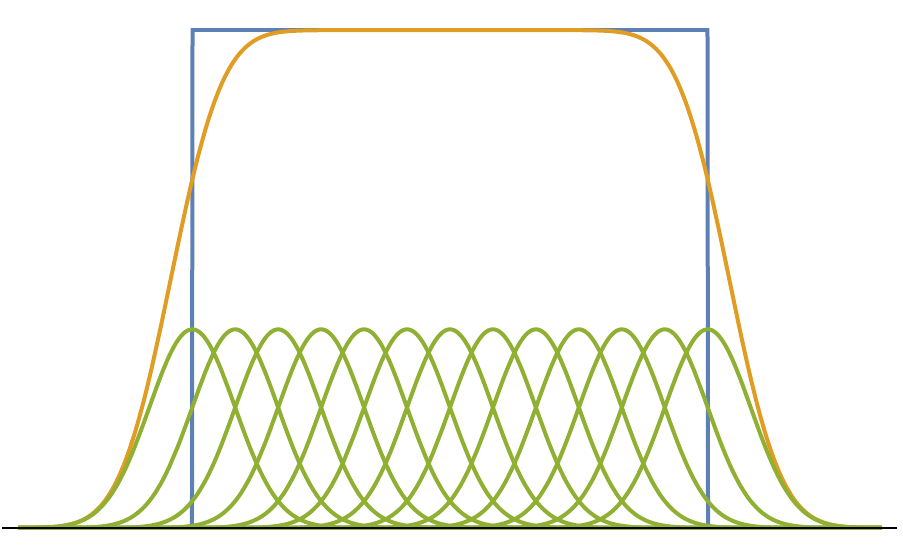}
	\caption{The desired boxcar function with width $2\sigma_y$ representing 
	the single slit and its approximation by a sum \eqref{boxc} of 13 shifted 
	Gaussians.}
	\label{fig:approximation}
\end{figure}

To model the sharp edge of the slit, we would ideally use a boxcar 
function.
However, although an analytical solution for the Schr\"odinger equation 
exists with these initial conditions, making use of the imaginary error 
function, the high-frequency components of its derivative pose numerical 
difficulties.
 Instead we can make a decent approximation  by taking  the $y$-component in \eqref{ss} to be the sum
\begin{equation} \psi^S(t,y) = \mathcal{N} \sum_{\ell=-K}^{K} \psi_y\left(t, 
y-\frac{\sigma_y}{K} \ell\right)\label{boxc}
\end{equation}
where $\mathcal{N}$ is an irrelevant normalization factor, and each $\psi_y$ is 
of the form \eqref{ga} with $v_0=0$ and with standard deviation 
$\sigma_y/\sqrt{2}K$.  For this diffraction experiment we take a square slit, 
assuming $\sigma_y = \sigma_z = 20 \hbar/mc$ is the half width of the slit; see 
Fig.~\ref{fig:approximation}. For our simulations we take $K=6$. Furthermore, 
knowing that the electron velocity is $0.1 c$, the de Broglie relation tells us 
that its wavelength will be $20 \pi \hbar / mc$, which is comparable to the 
full width of the slit. Therefore the near field will extend over at most a few 
times this distance.

\begin{figure}
		\includegraphics[width=8cm]{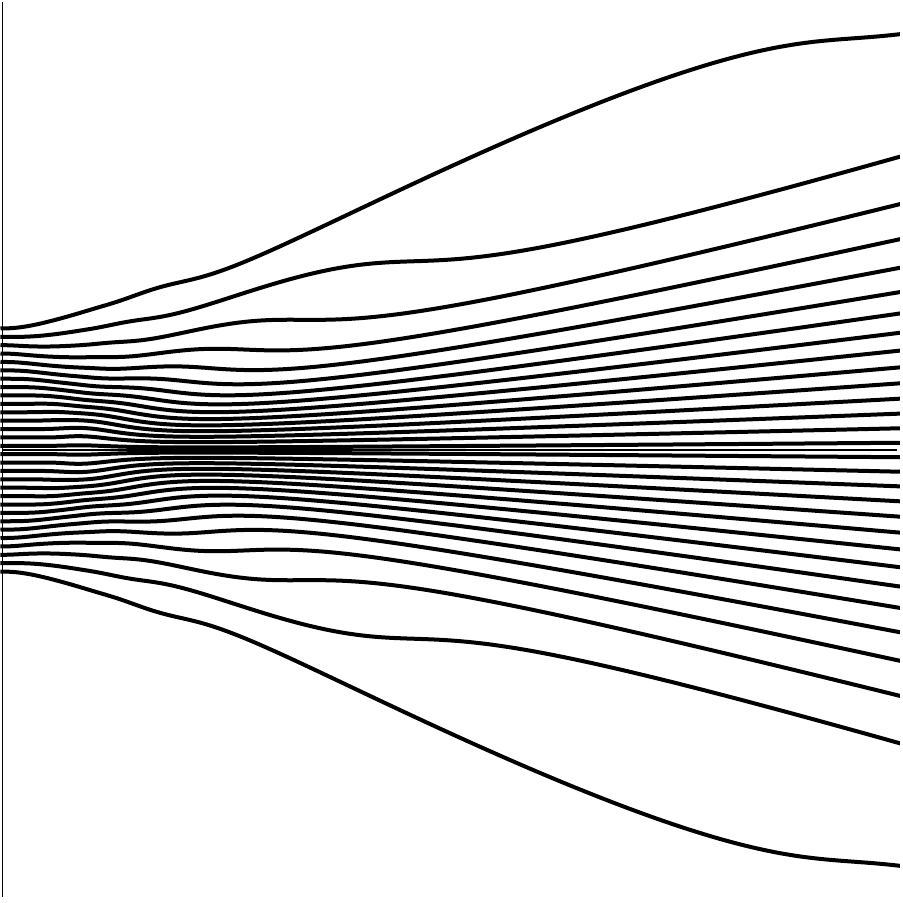}	\caption{Near-field 
		behavior of the trajectories for a single slit, without the curling and tumbling (keeping only the first term in \eqref{vf}).}
\label{nearf}
\end{figure}

We are ready to apply \eqref{vf}--\eqref{ta} in \eqref{gui}. To start, in 
Fig.~\ref{nearf} we give the so called Bohmian trajectories, i.e. the 
trajectories without including the curling and tumbling.  We plotted there the 
near-field behavior in the sense that the horizontal dimension of the picture 
covers only $100 \hbar / m c$.

\begin{figure}
	\begin{subfigure}{0.45\textwidth}
		\includegraphics[width=\textwidth]{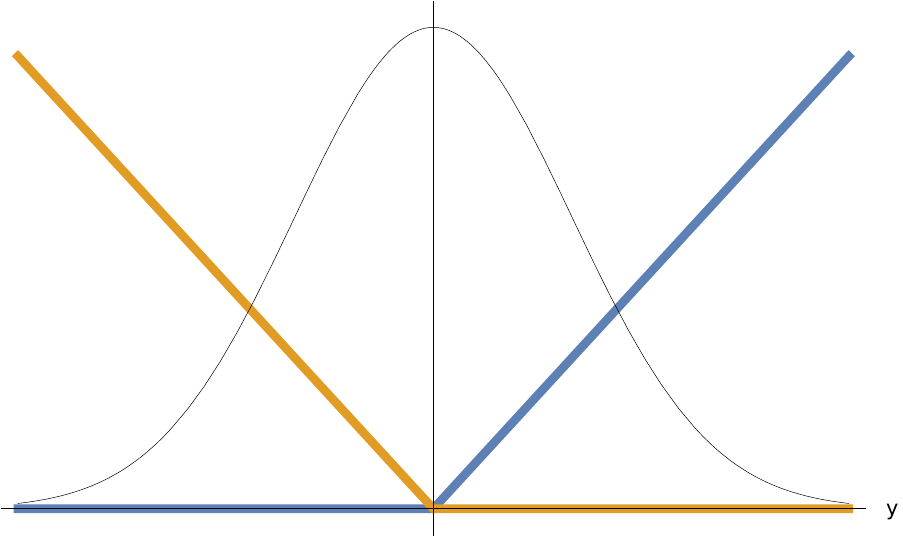}
		\caption{The tumbling rate for a single slit Gaussian wave function is linear 
			in $y$.}
		\label{fig:singleslittrajectoriesy}
	\end{subfigure}\hfill%
	\begin{subfigure}{0.45\textwidth}
		\includegraphics[width=\textwidth]{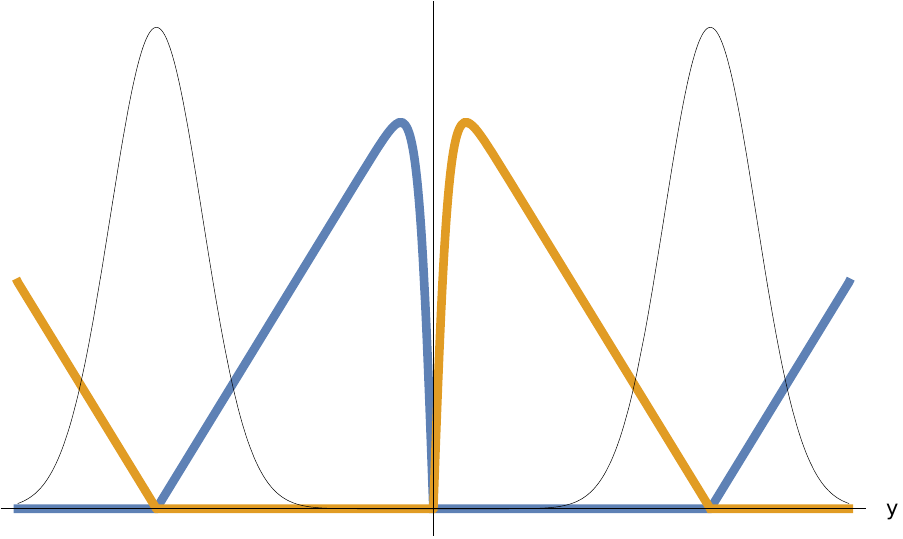}
		\caption{Right after exiting the double slits.}
		\label{fig:ratestart}
	\end{subfigure}
	\begin{subfigure}{0.45\textwidth}
		\includegraphics[width=\textwidth]{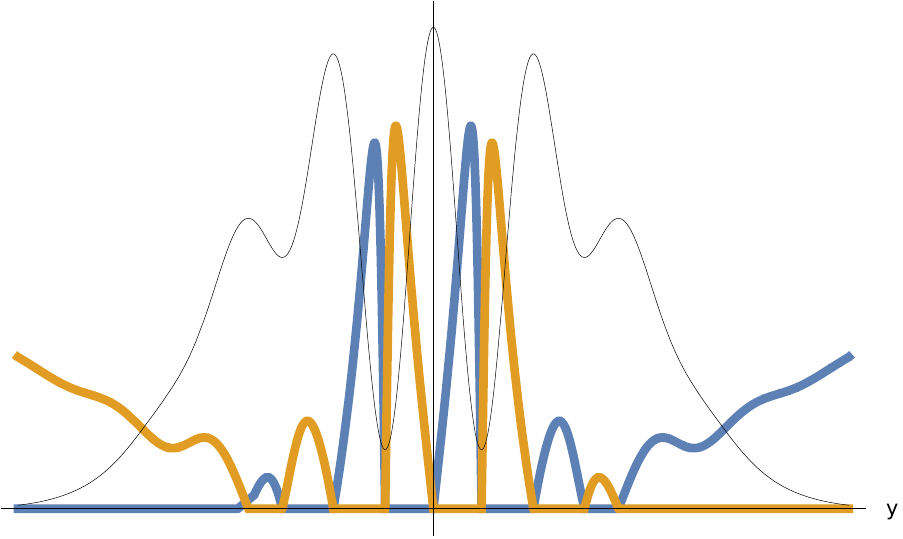}
		\caption{During formation of interference pattern.}
	\end{subfigure}\hfill%
	\begin{subfigure}{0.45\textwidth}
		\includegraphics[width=\textwidth]{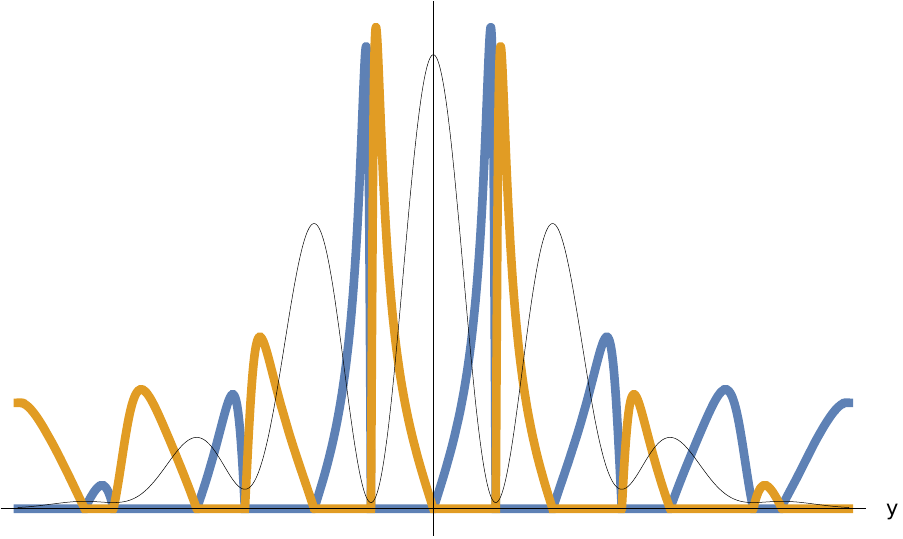}
		\caption{After the pattern has formed.}
	\end{subfigure}
	\caption{The $y$-dependence of the tumbling rates when $\vu{s}=\vu{y}$. The 
		blue curve corresponds to the tumbling rate from $+$ to $-$ chirality, and vice 
		versa for the orange curve. In black is shown $|\psi|^2$. For a single slit, 
		the profile is linear as shown in (a). For two slits, (b)-(d), oscillations 
		appear as the wave
		function starts to show more peaks farther away from the wall.}
	\label{fig:ratefunctions}
\end{figure}

An $(xy)$-projection of the trajectories including all terms in \eqref{vf}  is 
shown in Figs.~\ref{fig:singleslittrajectories} (a),(b) and (c) for different 
choices of spin orientation $\vu{s}$.  We also compare there the far-field 
behavior (Fraunhofer diffraction) with the near-field behavior (Fresnel 
diffraction).  The left panels in Fig.~\ref{fig:singleslittrajectories} show 
the far-field behavior with the screen at a distance $1000 \hbar / m c$ from 
the slit. The right panels give the near-field trajectories at one tenth of 
this distance. The left vertical line is the placement of the wall (in the 
$y$-direction). Because of the spreading of the wave function, the slits are 
not always visible on the largest scale. On the right, the location of the 
screen is indicated with a vertical line. Each time we show 25 trajectories for 
the far field and 5-10 trajectories for the near field, started from sampling 
initial positions according to the Born rule. Note that some trajectories cross 
the barrier of the slits. That is an artifact of our modeling due to the wave 
function having tails in these regions. For a wave function that vanishes in 
these regions, there would be no such crossings.  The particles tend to explore 
the support of the wave function, especially in the direction of the spin, in 
contrast to the trajectories without curling and tumbling, cf.\ 
Fig.~\ref{fig:singleslittrajectoriesd}. The particles will tend to keep moving 
at approximately the light speed in a certain direction until they reach a 
region of low density $|\psi|^2$. As is clear from the formulae 
\eqref{eqn:rate}--\eqref{ta}, in such regions the jump rate increases, giving 
more chance for a velocity flip to occur. This is also clarified through 
Fig.~\ref{fig:singleslittrajectoriesy} where the tumbling rate is plotted for 
$\psi_y$ a Gaussian of the form \eqref{ga} with $v_0=0$.

\begin{figure}
\begin{subfigure}{\textwidth}
\includegraphics[width=0.48\textwidth]{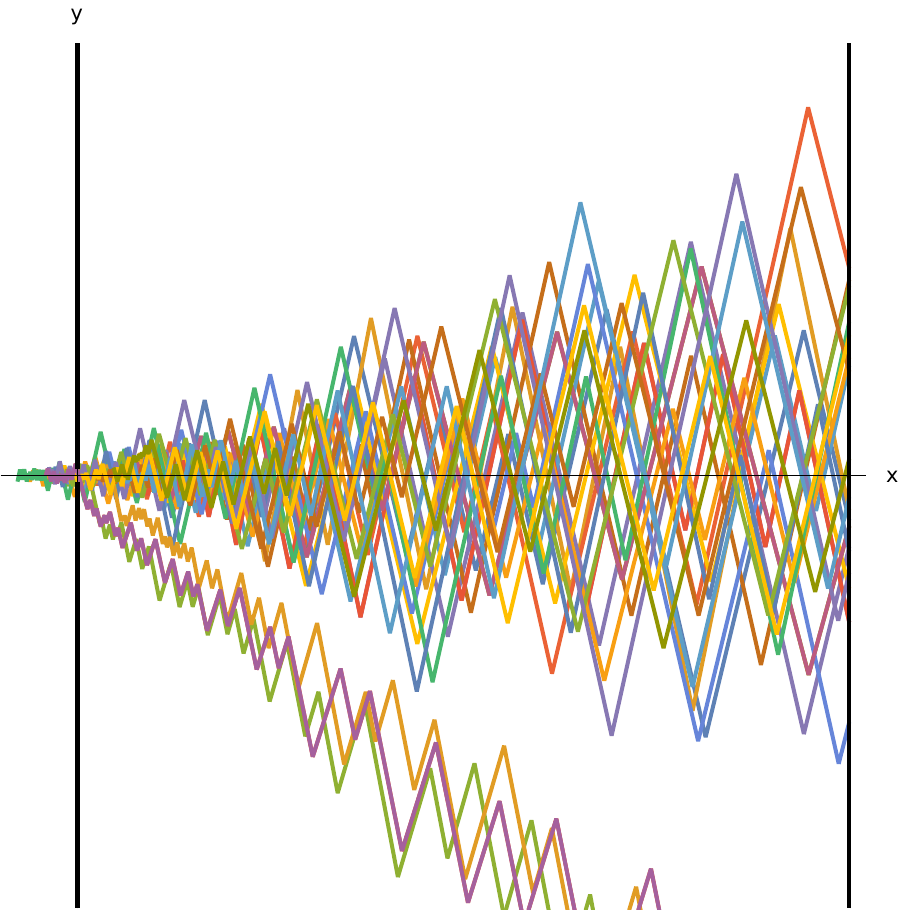}\hfill%
\includegraphics[width=0.48\textwidth]{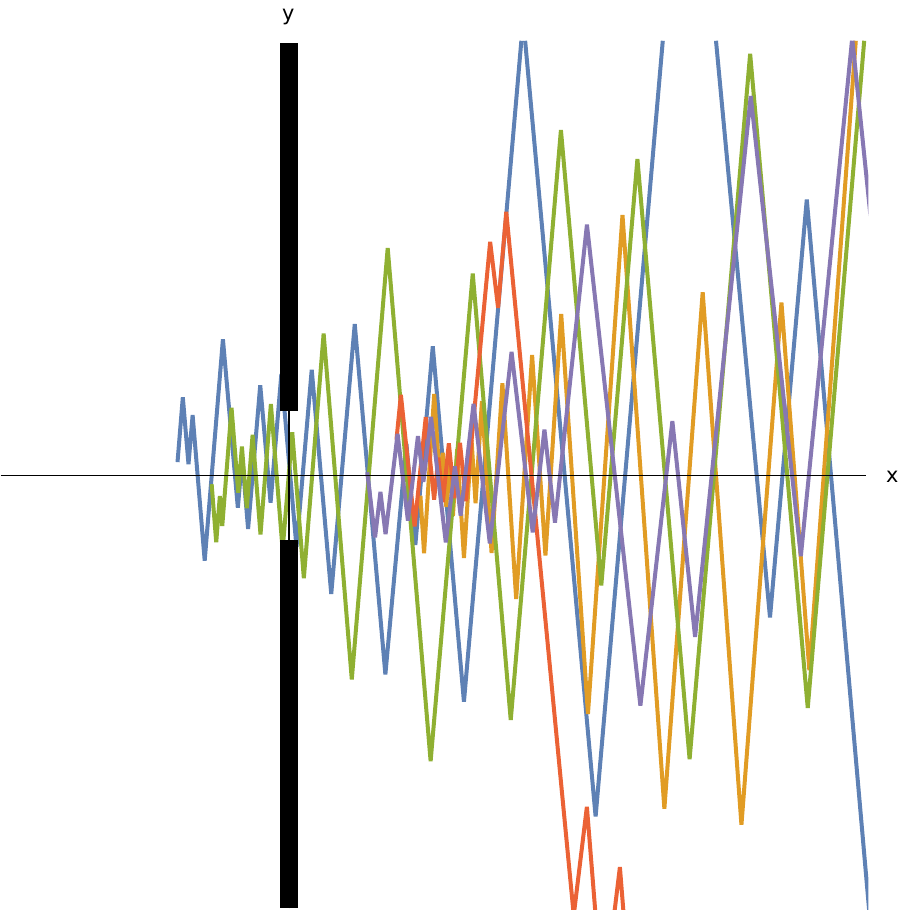}
\caption{$\vu{s} = \vu{y}$}
\end{subfigure}
\begin{subfigure}{\textwidth}
\includegraphics[width=0.48\textwidth]{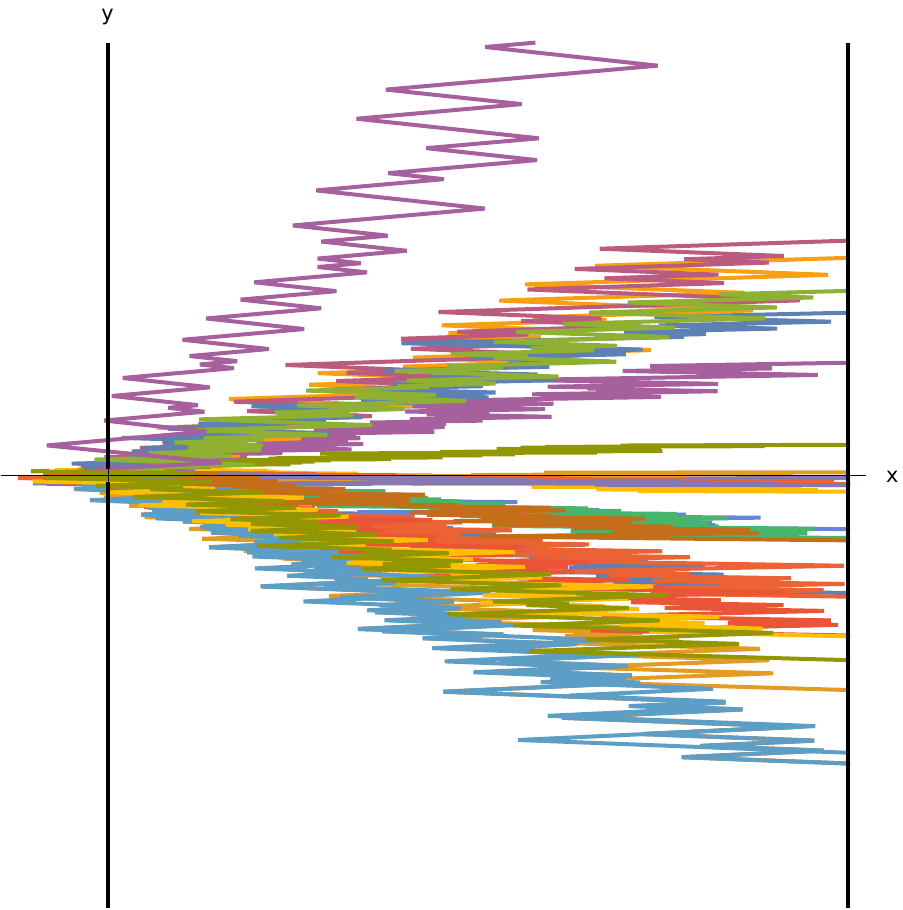}\hfill%
\includegraphics[width=0.48\textwidth]{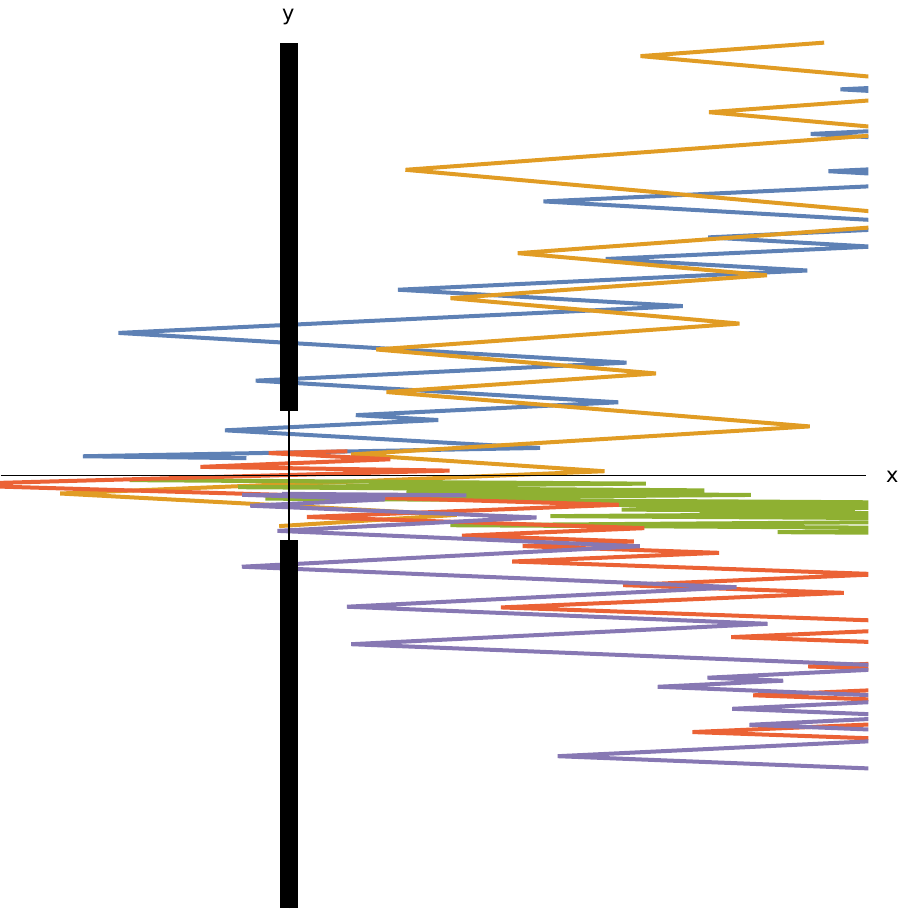}
\caption{$\vu{s} = \vu{x}$}
\end{subfigure}
\caption{Single slit trajectories with random initial positions distributed according to the Born rule for the  
initial wave function \eqref{ss} with the boxcar function \eqref{boxc} in the $y$-direction. Pictured on the left and 
the right are respectively the far-field and near-field behaviors.}
\label{fig:singleslittrajectories}
\end{figure}
\begin{figure}
\ContinuedFloat
\begin{subfigure}{\textwidth}
\includegraphics[width=0.48\textwidth]{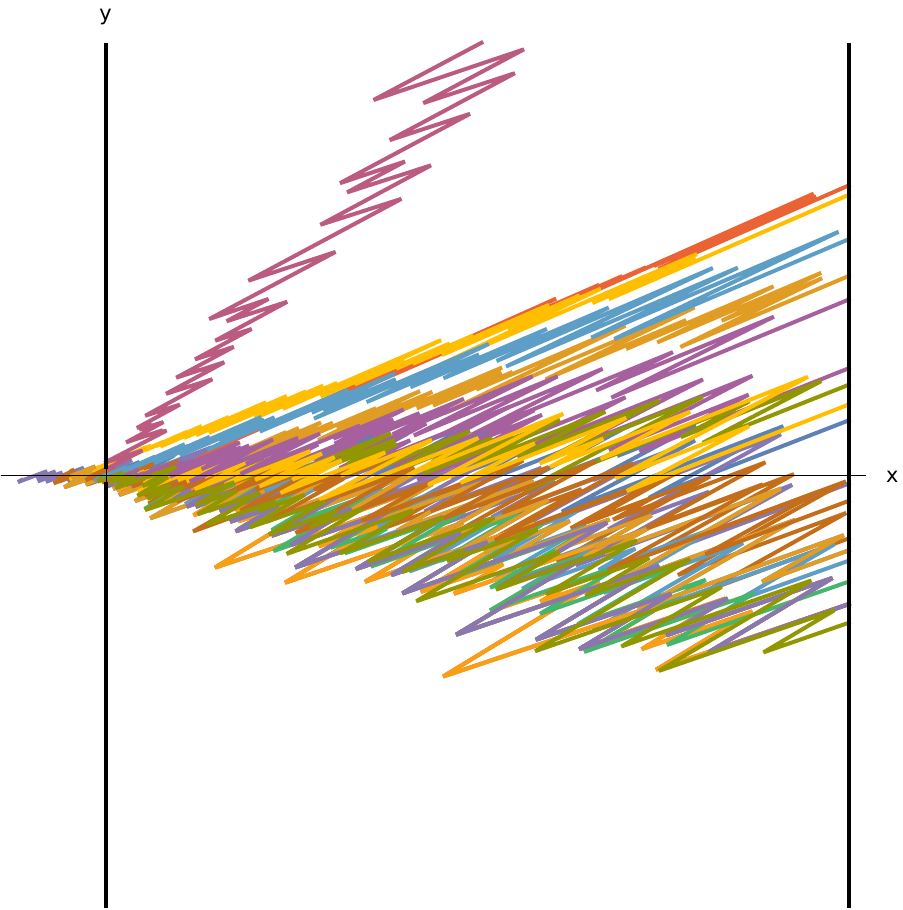}\hfill%
\includegraphics[width=0.48\textwidth]{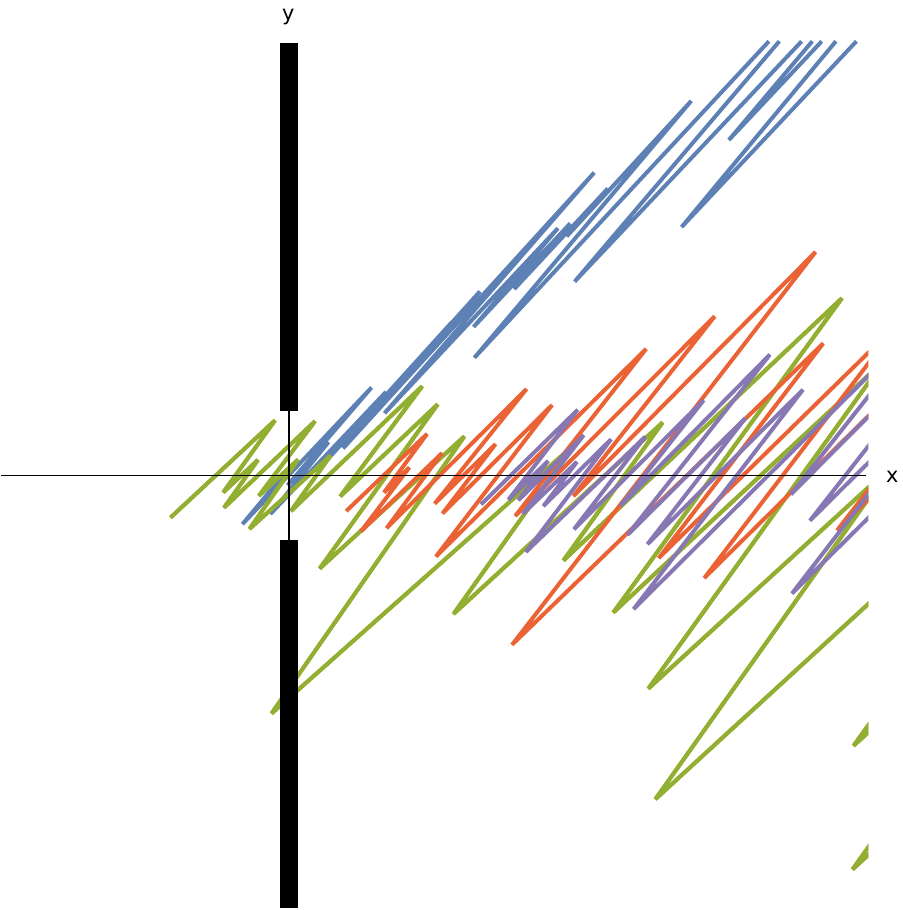}
\caption{$\vu{s} = (\vu{x}+\vu{y}+\vu{z})/\sqrt{3}$}
\end{subfigure}
\begin{subfigure}{\textwidth}
\includegraphics[width=0.48\textwidth]{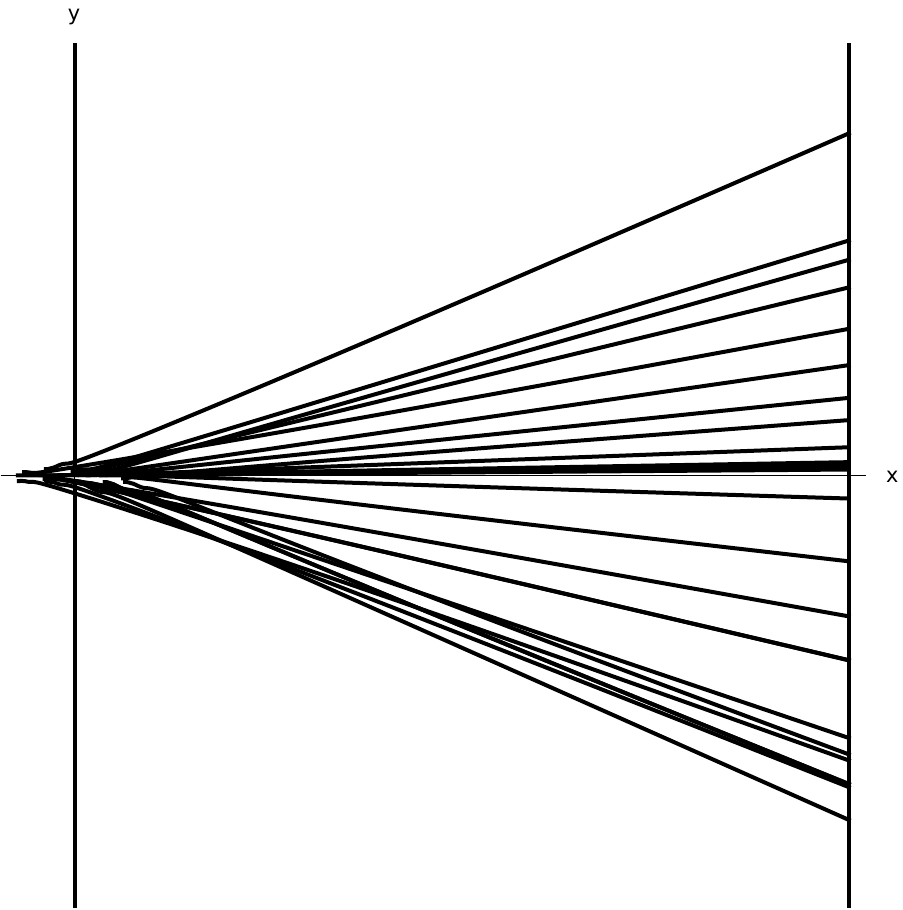}\hfill%
\includegraphics[width=0.48\textwidth]{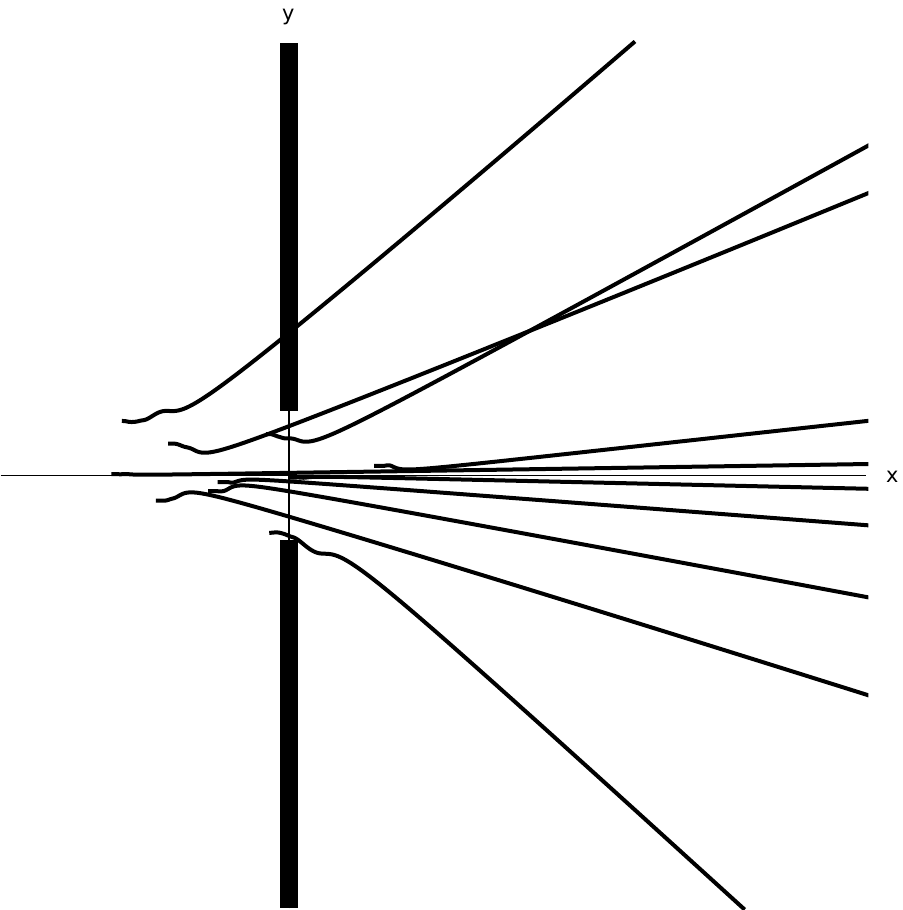}
\caption{Trajectories without the curling and tumbling (keeping only the first term in \eqref{vf}).\label{fig:singleslittrajectoriesd}}
\end{subfigure}
\caption{(continued)}
\end{figure}

\begin{figure}
	\begin{subfigure}{\textwidth}
		\includegraphics[width=0.48\textwidth]{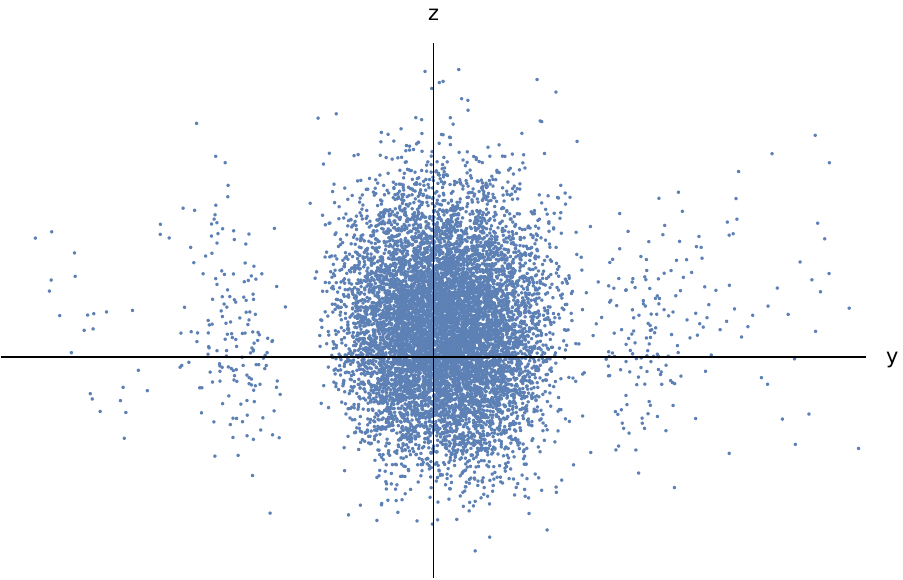}
		\includegraphics[width=0.48\textwidth]{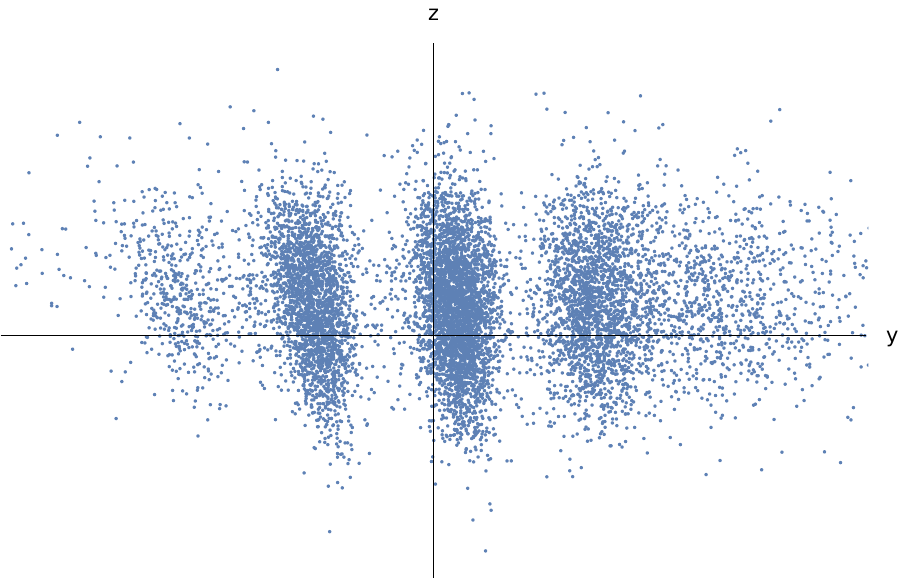}
		\caption{The two-dimensional pattern as it appears on the screen, 
		corresponding to the $y$- and $z$-coordinates of the particle.}
	\label{fig:pattern}
	\end{subfigure}
	\begin{subfigure}{\textwidth}
		\includegraphics[width=0.48\textwidth]{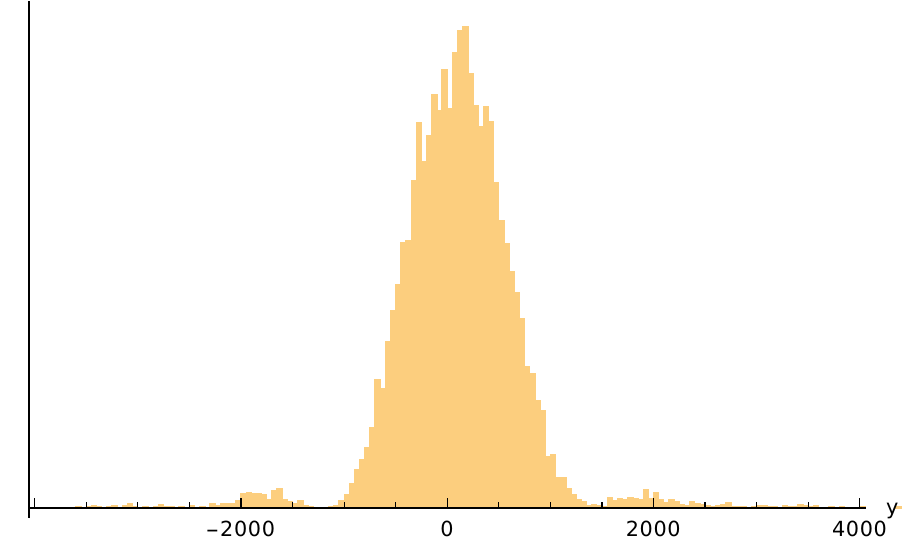}
	\includegraphics[width=0.48\textwidth]{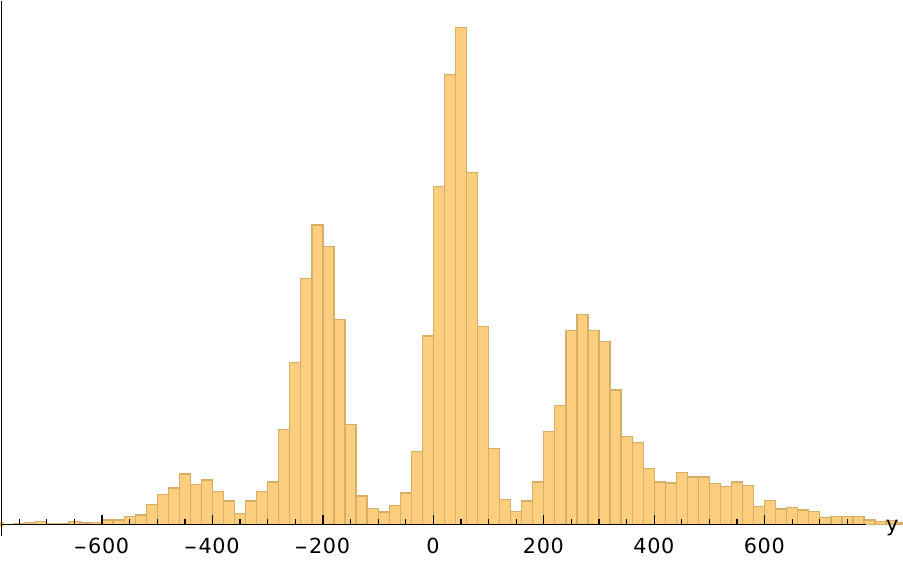}
		\caption{Distribution of the $y$-coordinate of the particle impacts on the 
		screen.}
	\label{fig:histogram}
	\end{subfigure}
	\begin{subfigure}{\textwidth}
		\includegraphics[width=0.48\textwidth]{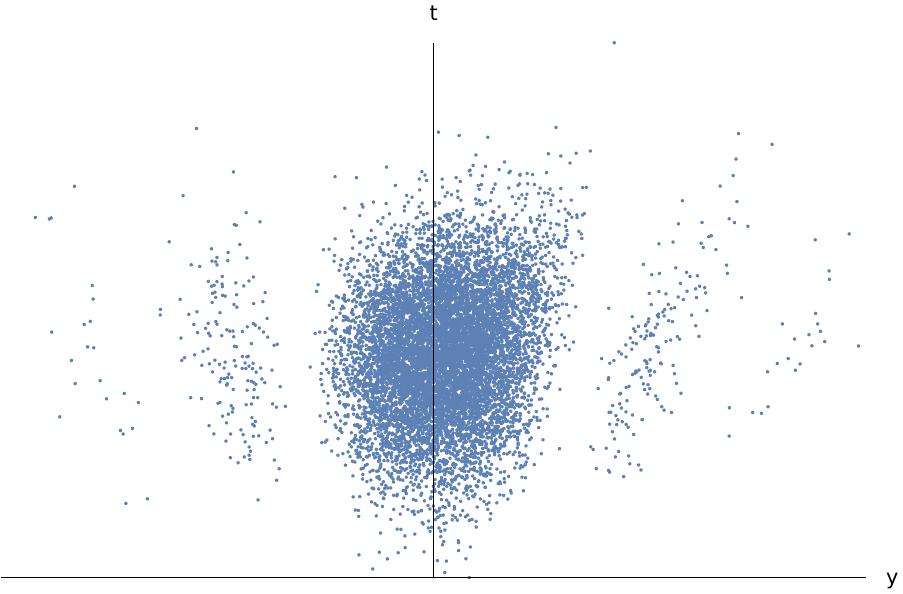}
		\includegraphics[width=0.48\textwidth]{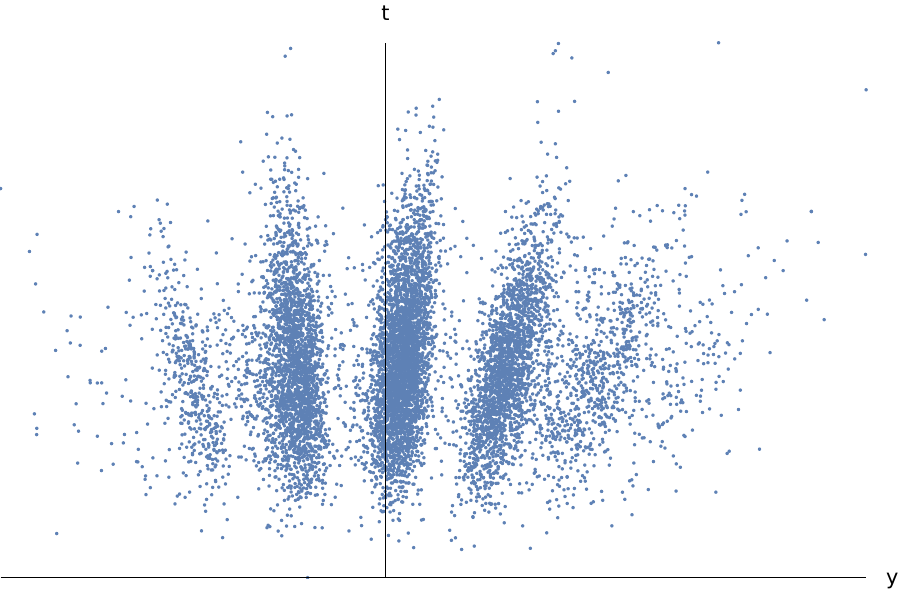}
\caption{Distribution of the $y$-coordinate of the particle impacts in function 
of the arrival time.}
	\label{fig:evolution}
	\end{subfigure}
	\caption{Patterns on the screen, for $\vu{s} = 
	(\vu{x}+\vu{y}+\vu{z})/\sqrt{3}$. The single slit is on the left and the double slit is on the right. }
	\label{fig:singleslitscreen}
	\label{fig:doubleslitscreen}
\end{figure}

Taking the spin to point towards $(\vu{x}+\vu{y}+\vu{z})/\sqrt{3}$, the 
distribution of the positions at impact (first crossing) of the particles on the 
screen is shown in Fig.~\ref{fig:singleslitscreen}, on the left. Notice the secondary 
maxima, characteristic of diffraction. This distribution is not just simply equal to $\abs{\psi}^2$. First of all, the wave function changes over time and some trajectories may take more time than others to reach the screen. Second, it is also the first crossing of the screen that is recorded; due to the zig-zag a trajectory may cross the screen several times. (We have here ignored what exactly the effect of a physical screen is on the trajectories.)

In Fig.~\ref{fig:evolution}, on the left, the distribution of the $y$ coordinate is plotted in function of the arrival time, which clearly illustrates the time-dependence of the pattern.
The arrival time for each particle trajectory is the time when the particle first hits the screen (at a certain fixed distance from the slit(s)).   Note that the pattern, perhaps surprisingly, appears asymmetric around $y=0$, with offset 
towards higher $y$- and $z$-coordinates.  The origin of that asymmetry are particles hitting the screen 
whilst on an excursion in the direction of the spin vector, away from the bulk 
of the wave packet. That, in other words, is an effect due to the electron's spin direction, here visualized explicitly.  

\subsection{Double slit interference}

Moving on to the double slit experiment, $b$ denotes half the distance between 
the centers of the slits, $b=6\sigma_y = 120\hbar / mc$, and the height of the 
slits is $\sigma_z = 5\sigma_y$. As the focus now lies on the interference effects (and 
not diffraction\footnote{There are of course also diffraction gratings that involve many slits, and interference underlies diffraction phenomena.  The terminology of this paper wants foremost to distinguish between a single and double slits.}), for simplicity we use a superposition of single Gaussians for 
the $y$-component $\psi^S$ in \eqref{ss}:
\begin{equation}\label{super}
\psi^S(t,y) = \frac{1}{\sqrt{2}} \psi_y(t,y-b) + \frac{1}{\sqrt{2}} 
\psi_y(t,y+b)
\end{equation}
The $x$- and $z$-components are unchanged from the previous section, and in particular, it is mainly the de Broglie wavelength $\hbar/{mv_0}$ that determines the shape of the interference pattern.\\
For understanding the tumbling, we need e.g.\ the logarithmic derivative, entering \eqref{vf},
\begin{equation}\label{yder}
\frac{1}{\psi^G} \pdv{\psi^G}{y} = \frac{1}{2\sigma_y^2 (1 + i \gamma t) }\left(-y 
+ b \tanh{ \frac{by}{2\sigma_y^2(1+i\gamma t)}}\right)
\end{equation}
where $\gamma= \hbar/ 2 m \sigma_y^2 $.
The nonzero distance between the slits adds a term proportional to $b$. As is 
clear from \eqref{ta}, that is important in the tumbling rate when the spin $ 
\vu{s}$ has a component in the $y$-direction. For large times $t$, deviations 
from the
linear behavior of the rate function will be most noticeable 
near the minima of the interference pattern, as shown in Fig.~\ref{fig:ratefunctions}. Notice how the rate function right after the slits, 
shown in Fig.~\ref{fig:ratestart}, approximates a piecewise linear function, 
with zeroes near the local maxima. As the interference pattern forms, the rate 
function takes on a more complicated shape, peaking where the wave function is 
small, and approximating a linear function where the wave function is 
approximately Gaussian.

Investigating the patterns that appear on the screen, we again fix $\vu{s} = 
(\vu{x}+\vu{y}+\vu{z})/\sqrt{3}$. As was the case with diffraction, even though 
interference fringes are clearly visible, cf.\ Fig.~\ref{fig:doubleslitscreen}, they are not directly related to 
$\abs{\psi}^2$; see similar (and also experimental) results and considerations in
\cite{kurtsiefer1997,kellers2017}.  Fig.~\ref{fig:evolution} on the right strikingly resembles experimental results in \cite{kurtsiefer1997}.

\begin{figure}
\begin{subfigure}{\textwidth}
\includegraphics[width=0.48\textwidth]{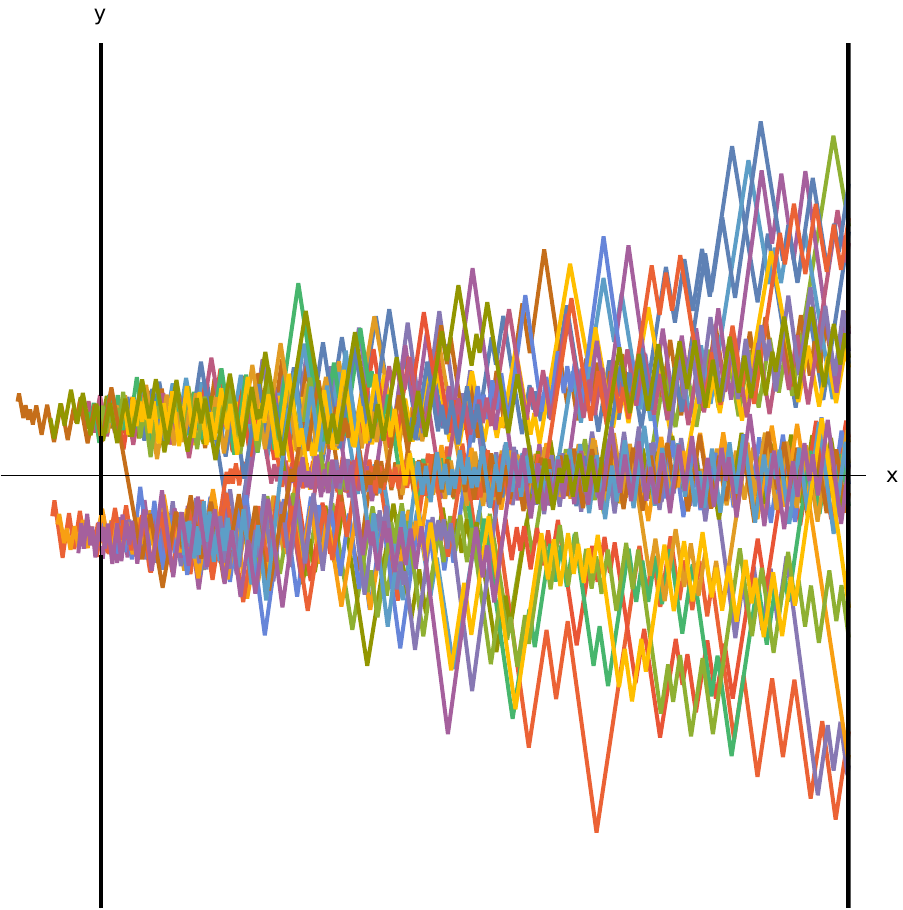}\hfill%
\includegraphics[width=0.48\textwidth]{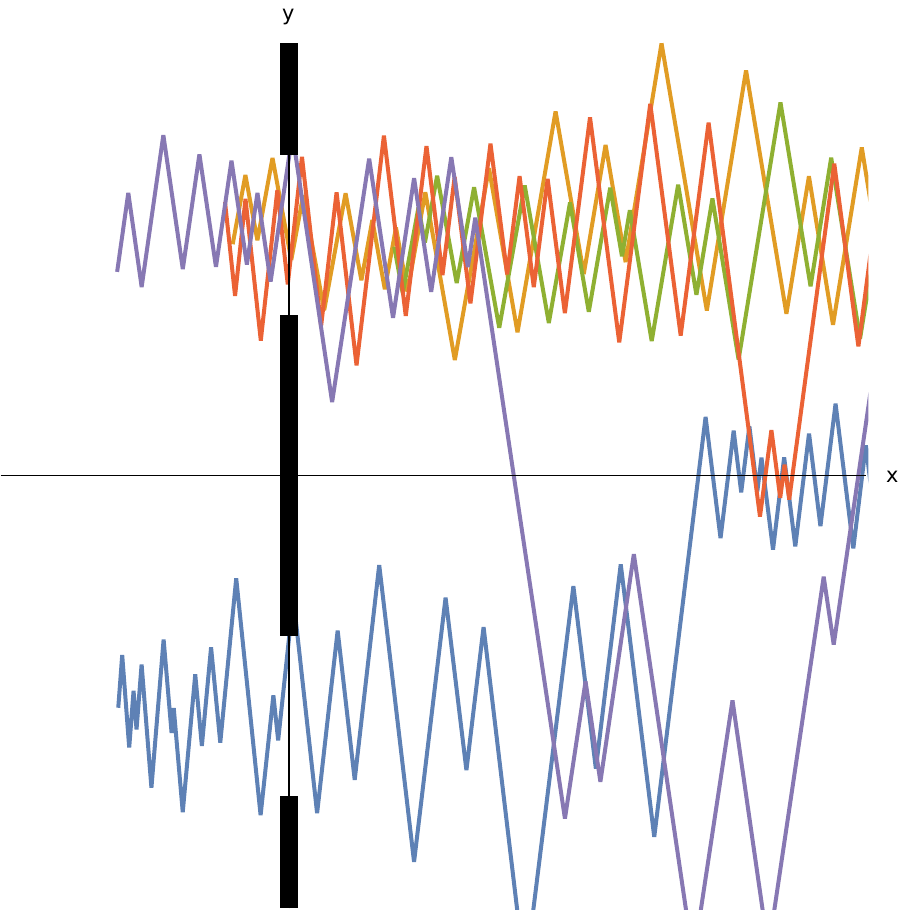}
\caption{$\vu{s} = \vu{y}$}
\end{subfigure}
\begin{subfigure}{\textwidth}
\includegraphics[width=0.48\textwidth]{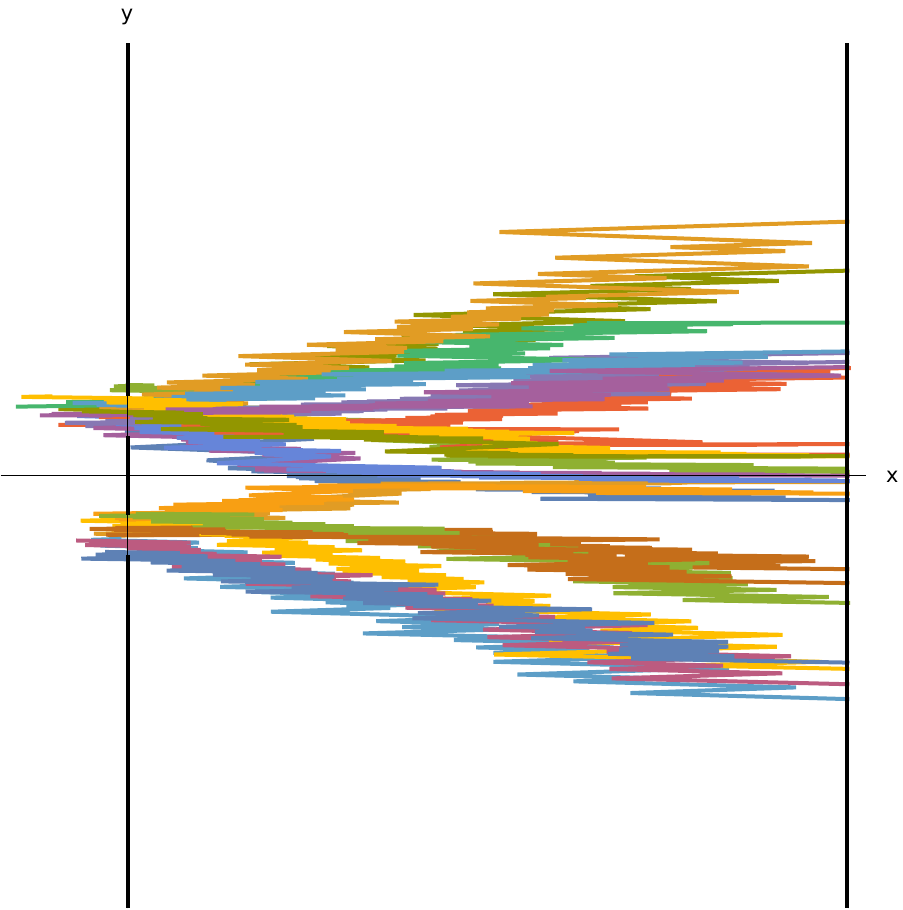}\hfill%
\includegraphics[width=0.48\textwidth]{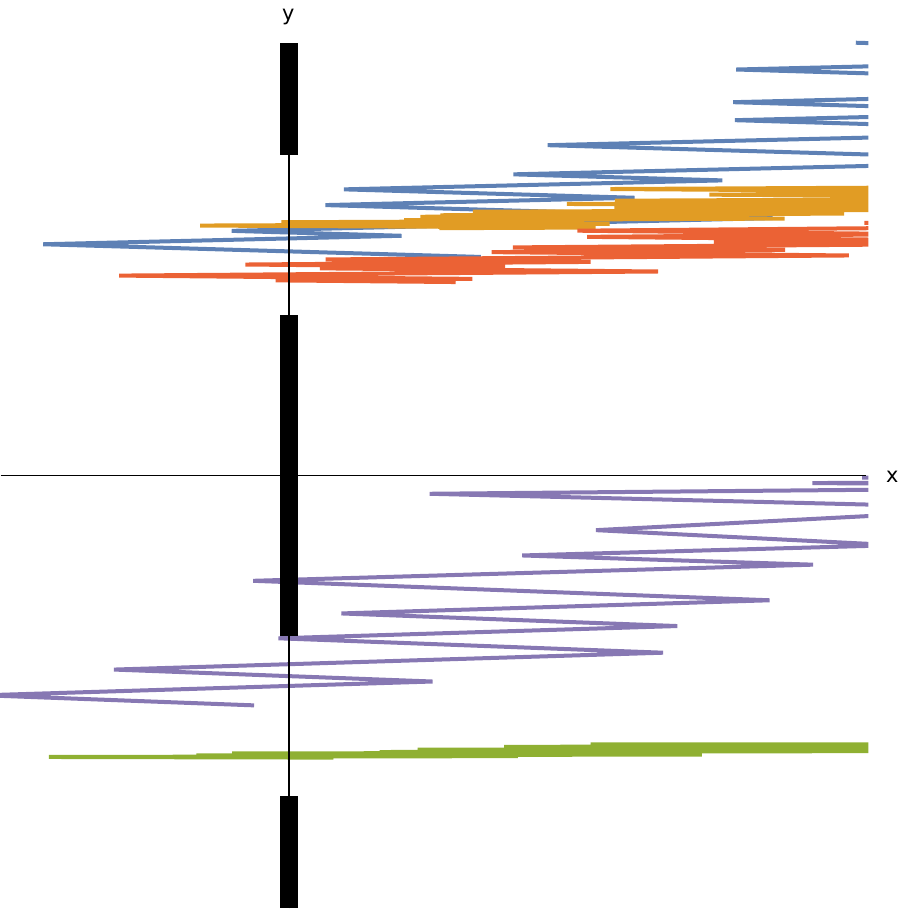}
\caption{$\vu{s} = \vu{x}$}
\end{subfigure}
\caption{Double slit trajectories with random initial positions distributed according to the Born rule for the  
initial wave function \eqref{ss} with the Gaussian superposition \eqref{super} in the $y$-direction. Pictured on the left and 
the right are respectively the far-field and near-field behaviors.}
\label{fig:doubleslittrajectories}
\end{figure}
\begin{figure}
\ContinuedFloat
\begin{subfigure}{\textwidth}
\includegraphics[width=0.48\textwidth]{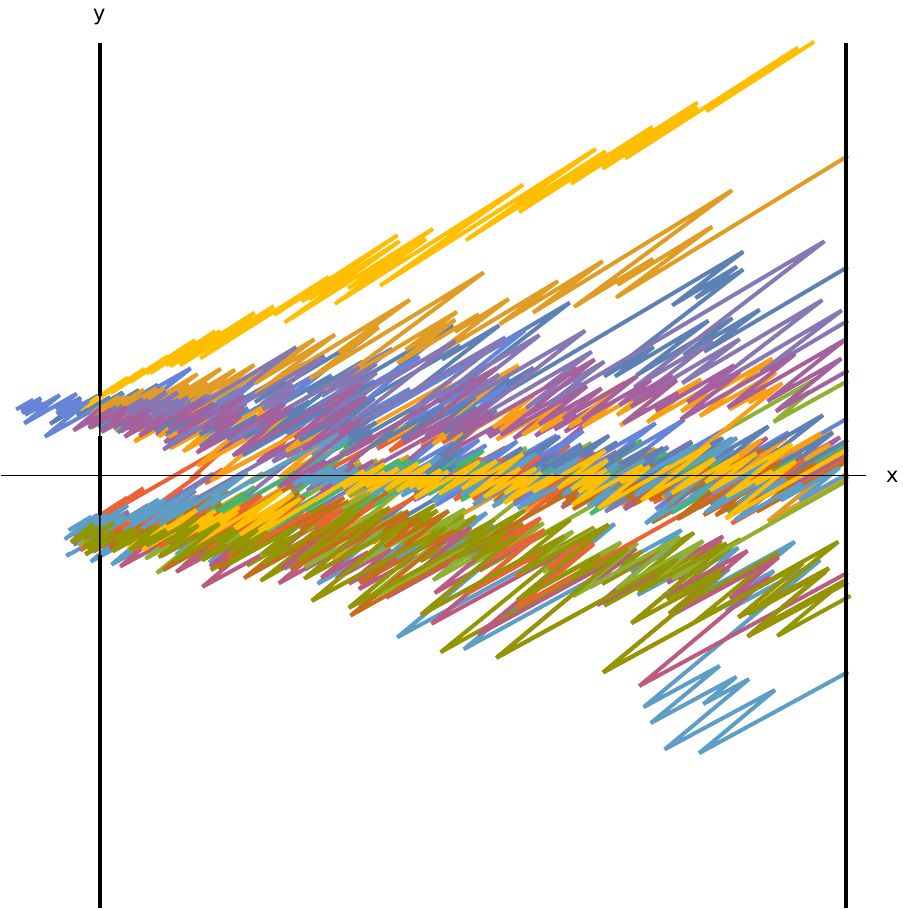}\hfill%
\includegraphics[width=0.48\textwidth]{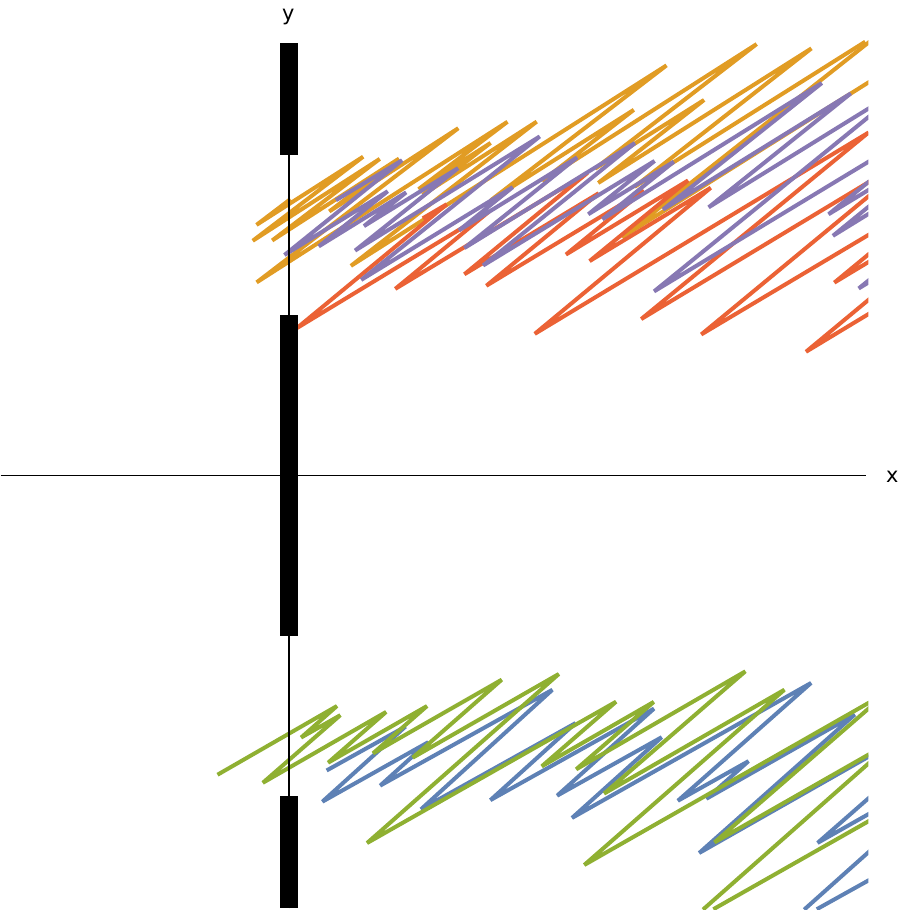}
\caption{$\vu{s} = (\vu{x}+\vu{y}+\vu{z})/\sqrt{3}$}
\end{subfigure}
\begin{subfigure}{\textwidth}
\includegraphics[width=0.48\textwidth]{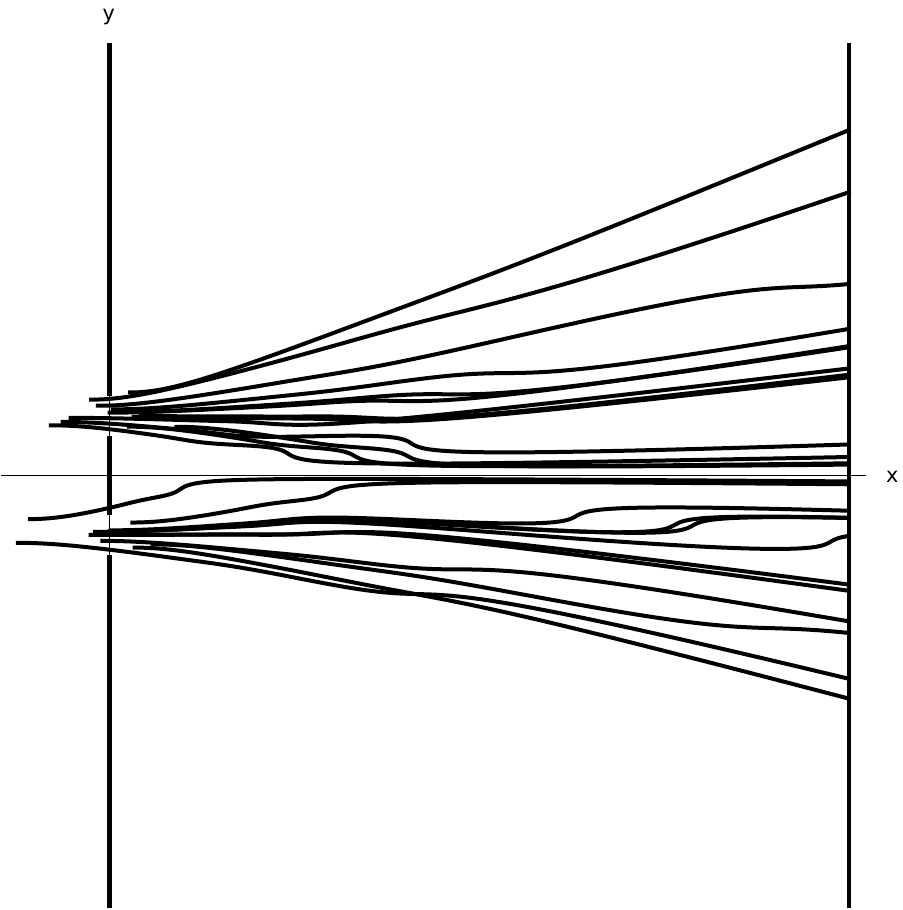}\hfill%
\includegraphics[width=0.48\textwidth]{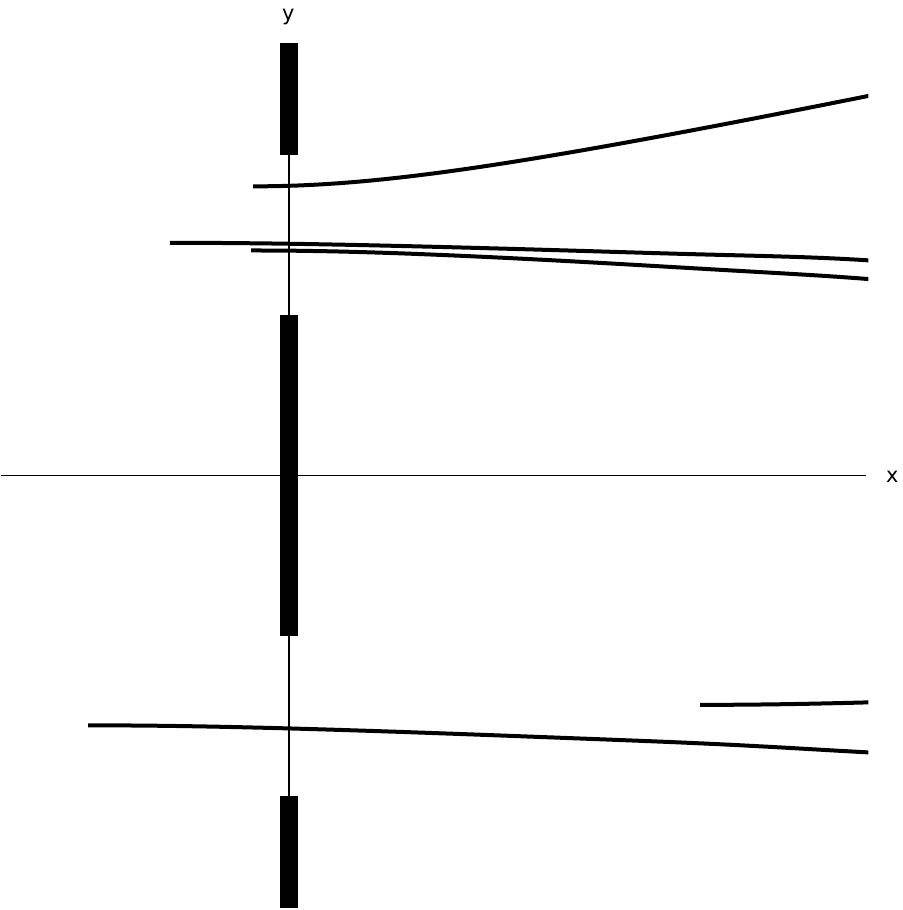}
\caption{Trajectories without curling and tumbling (only keeping the first term in \eqref{vf}).}
\end{subfigure}
\caption{(continued)}
\end{figure}

\begin{figure}
	\begin{subfigure}{0.45\textwidth}
		\includegraphics[width=\textwidth]{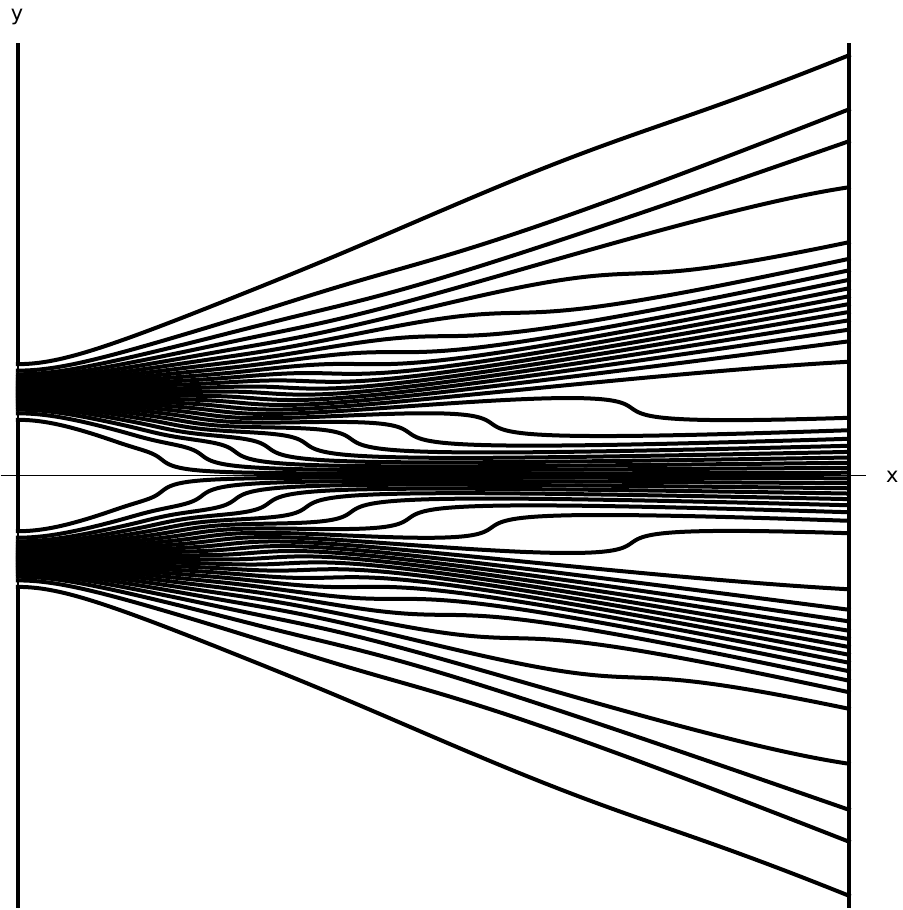}
		\caption{Trajectories without curling and tumbling (keeping only the first term in \eqref{vf}).\label{fig:doubleslittrajectories2-1}}
	\end{subfigure}\hfill%
	\begin{subfigure}{0.45\textwidth}
		\includegraphics[width=\textwidth]{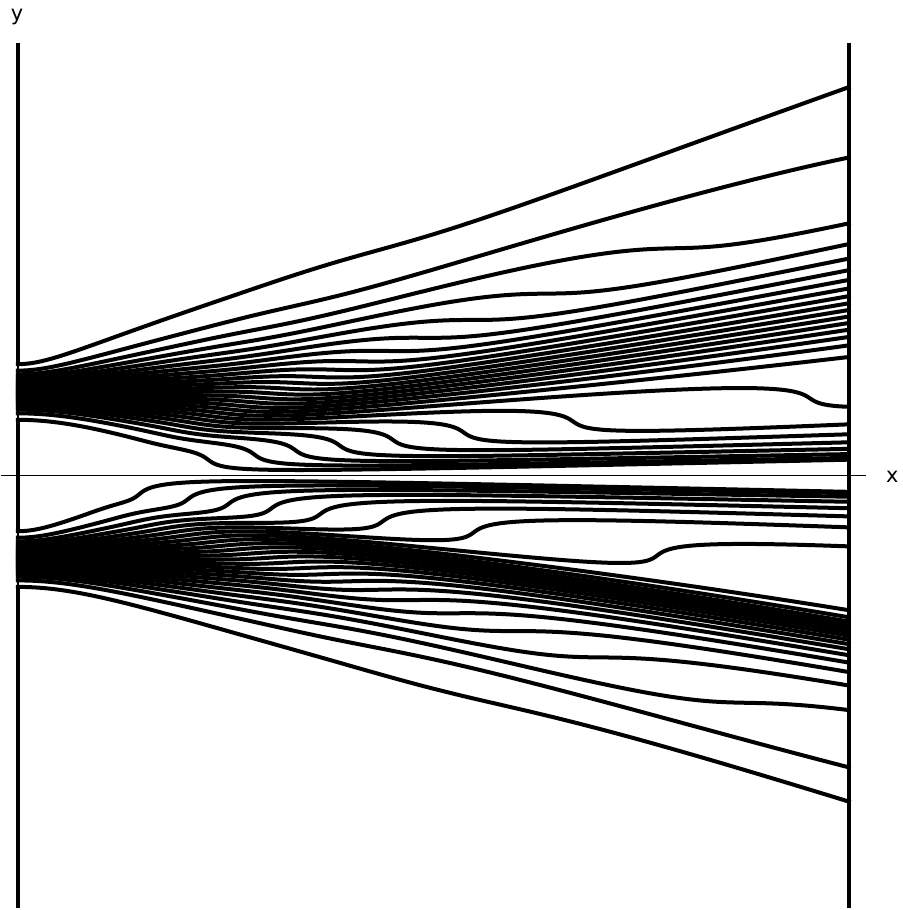}
		\caption{Trajectories without tumbling (without the third term in \eqref{vf}).\label{fig:doubleslittrajectories2-2}}
	\end{subfigure}
	\begin{subfigure}{0.45\textwidth}
		\includegraphics[width=\textwidth]{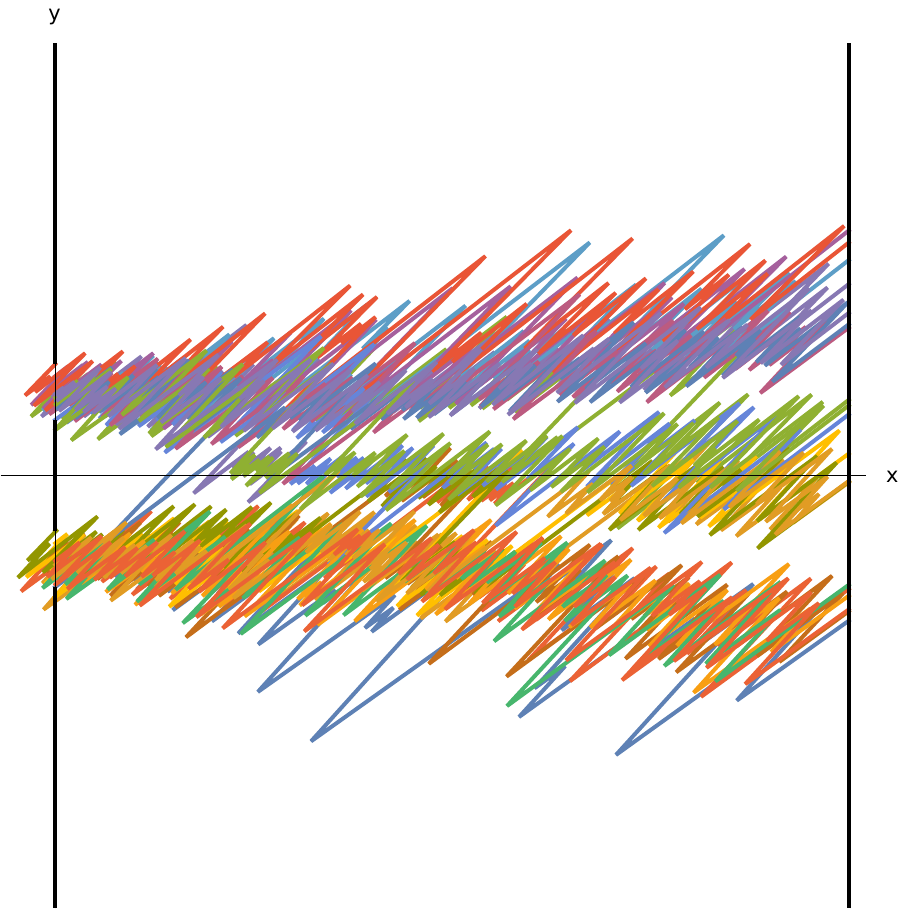}
		\caption{Trajectories with curling and tumbling.
		\label{fig:doubleslittrajectories2-3}
		}
	\end{subfigure}\hfill%
	\begin{subfigure}{0.45\textwidth}
		\includegraphics[width=\textwidth]{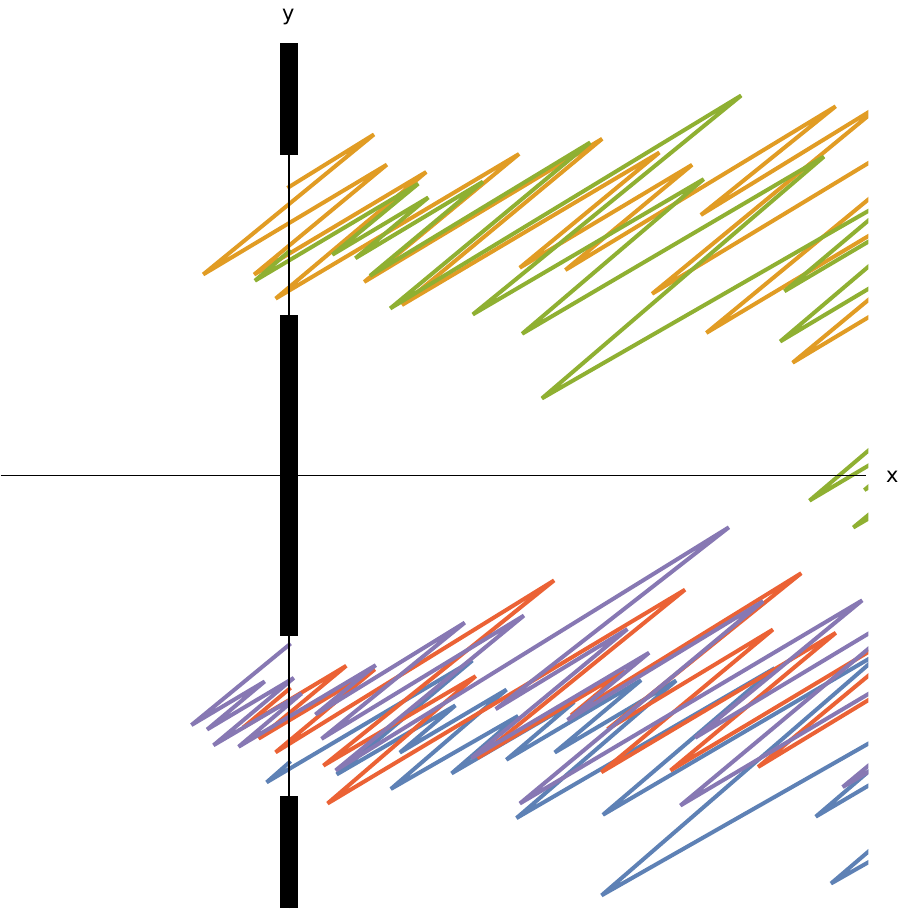}
		\caption{Near-field behavior of (c).}
	\end{subfigure}%
\caption{Double slit trajectories for particles starting at $x=0$ and $z=0$, with the same initial positions in (a)-(c), and $\vu{s} = (\vu{x}+\vu{y}+\vu{z})/\sqrt{3}$. Comparison of the different dynamics.}
	\label{fig:doubleslittrajectories2}
\end{figure}

\begin{figure}
	\begin{subfigure}{0.45\textwidth}
		\includegraphics[width=\textwidth]{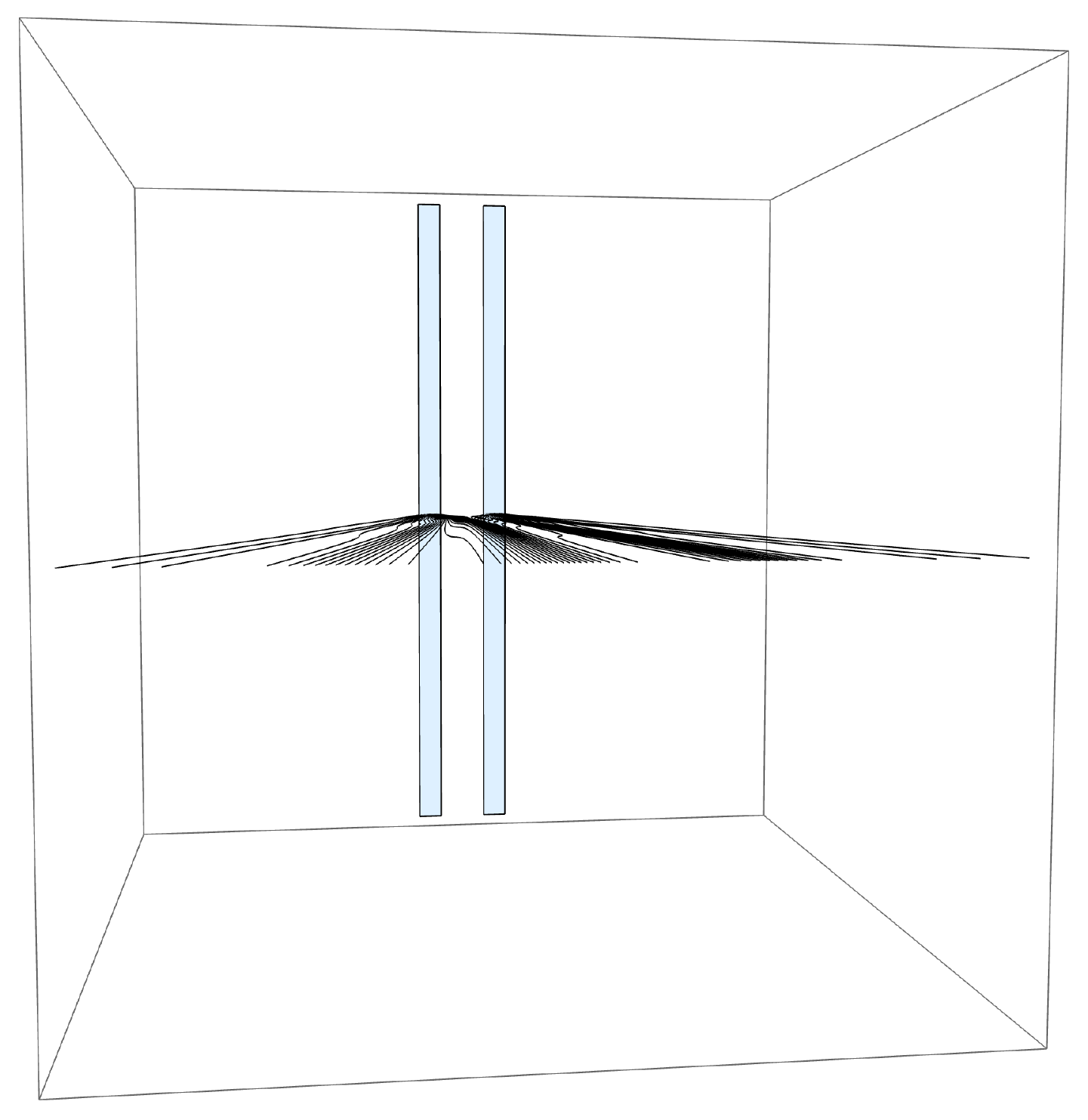}
		\caption{Without tumbling and curling.}
	\end{subfigure}\hfill%
	\begin{subfigure}{0.45\textwidth}
		\includegraphics[width=\textwidth]{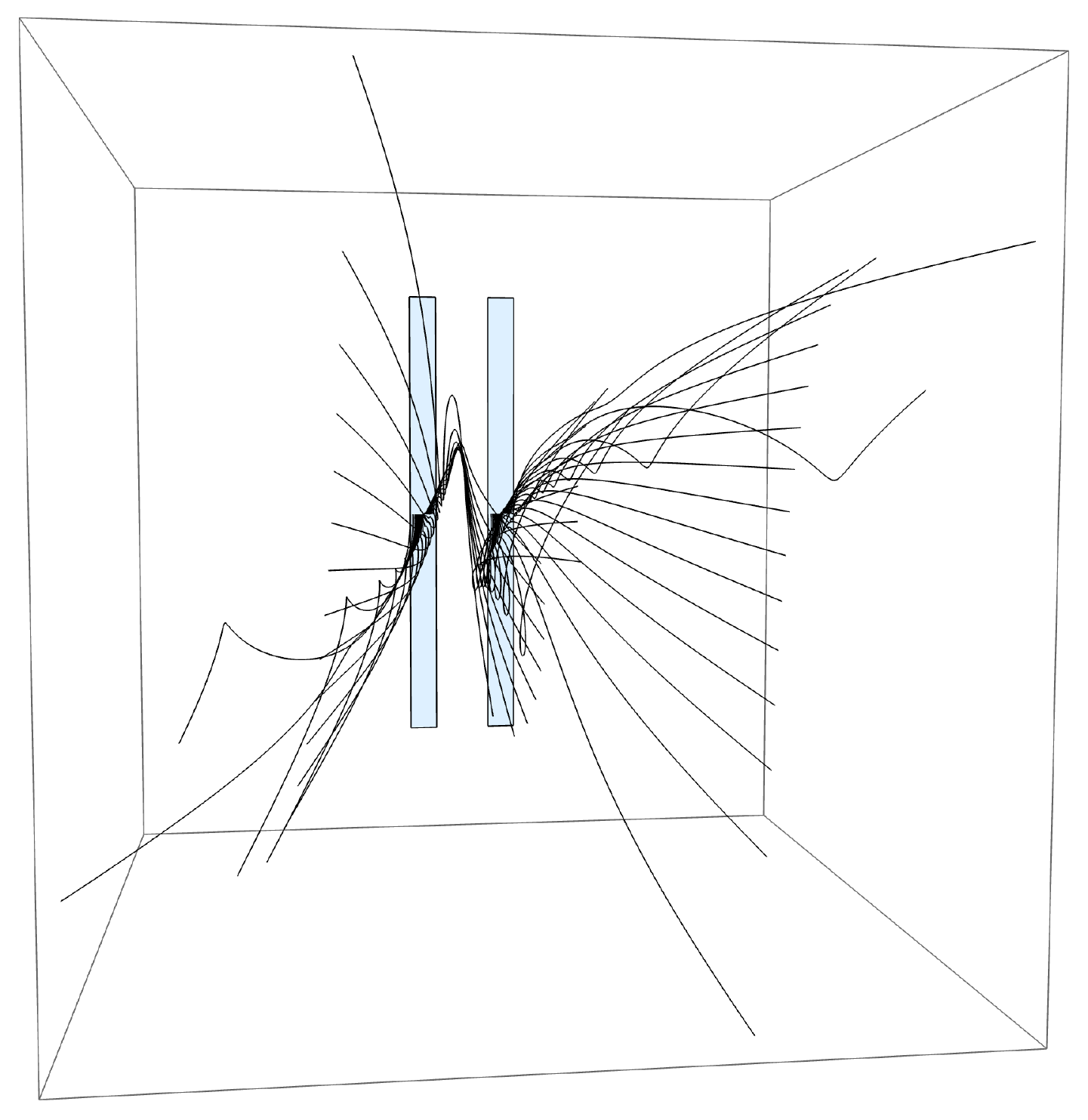}
		\caption{Without tumbling, only curling, $\vu{s} = \vu{x}$.}
	\end{subfigure}
	\begin{subfigure}{0.45\textwidth}
		\includegraphics[width=\textwidth]{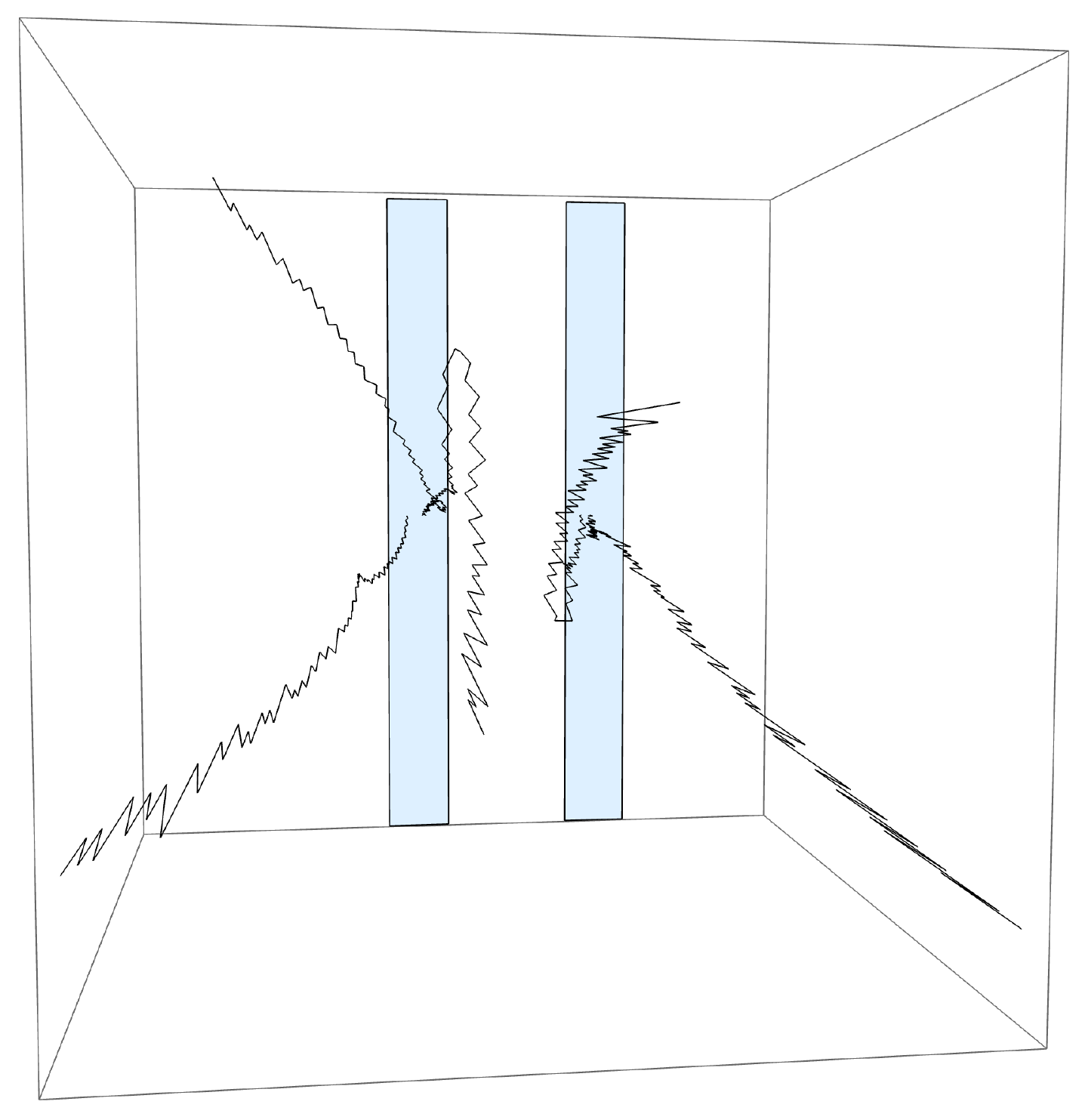}
		\caption{With curling and tumbling, $\vu{s} = \vu{x}$.}
	\end{subfigure}\hfill%
	\begin{subfigure}{0.45\textwidth}
		\includegraphics[width=\textwidth]{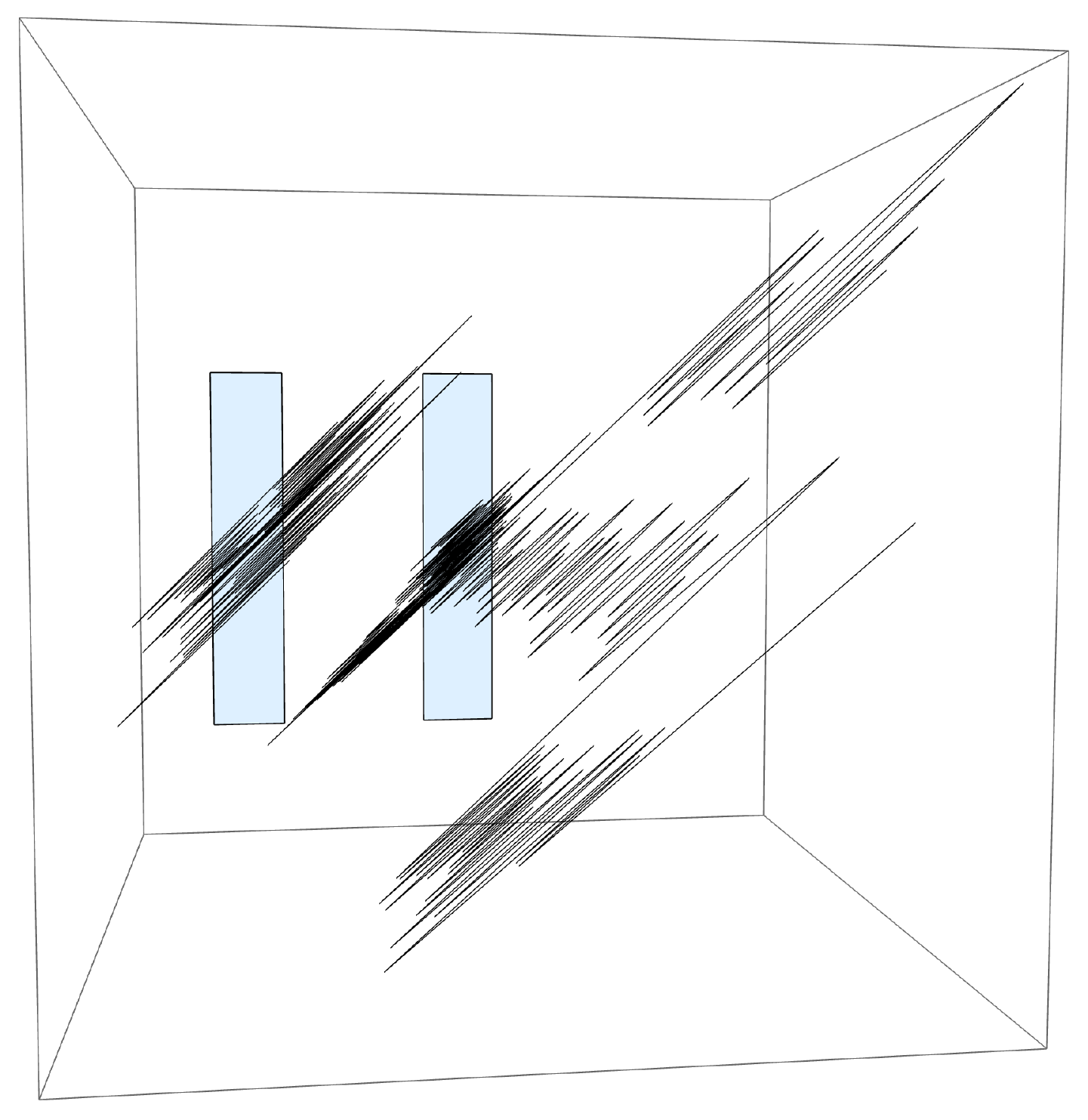}
		\caption{With curling and tumbling, $\vu{s} = (\vu{x}+\vu{y}+\vu{z})/\sqrt{3}$.}
	\end{subfigure}
\caption{Three-dimensional views towards wall.  Particles start from $x=0$ and $z=0$.}
	\label{fig:doubleslittrajectories3}
\end{figure}

The trajectories for various directions of the spin are plotted in Fig.~\ref{fig:doubleslittrajectories}. A comparison can now be made of the different dynamics, obtained by the various contributions in the velocity field \eqref{vf}. For the double slit experiment, trajectories without the curling and tumbling were pioneered in 
\cite{philippidis1} and without tumbling in  \cite{philippidis2}. 
See Fig.~\ref{fig:doubleslittrajectories2} for the view of the 
trajectories in the $(xy)$-plane and Fig.~\ref{fig:doubleslittrajectories3} 
for three-dimensional views. Contrary to Fig.~\ref{fig:doubleslittrajectories} 
we do not adhere to the Born rule here, constraining the particles to start at 
$x=0$ and $z=0$, with the same initial positions in each figure.  Without the 
curling and tumbling, there is a symmetry around the $x$-axis, cf.\ 
Fig.~\ref{fig:doubleslittrajectories2-1}. This symmetry is slightly spoiled 
when curling is introduced, cf.\ Fig.~\ref{fig:doubleslittrajectories2-2} (as 
was already noted in \cite{philippidis2}), and even more so when the tumbling 
is included, cf.\ Fig.~\ref{fig:doubleslittrajectories2-3}. In the latter case, 
the tumbling also causes the particle to occasionally strays back to the other 
side of the wall. As the interference pattern develops, the probability 
distribution forms channels and the particle tends to get stuck in one of those 
channels. As mentioned before when we discussed diffraction, this is because 
the jump rate is such that the probability of reflection increases when the 
particle moves to a low density region.

\begin{figure}[h]
\begin{subfigure}{0.45\textwidth}
\includegraphics[width=\textwidth]{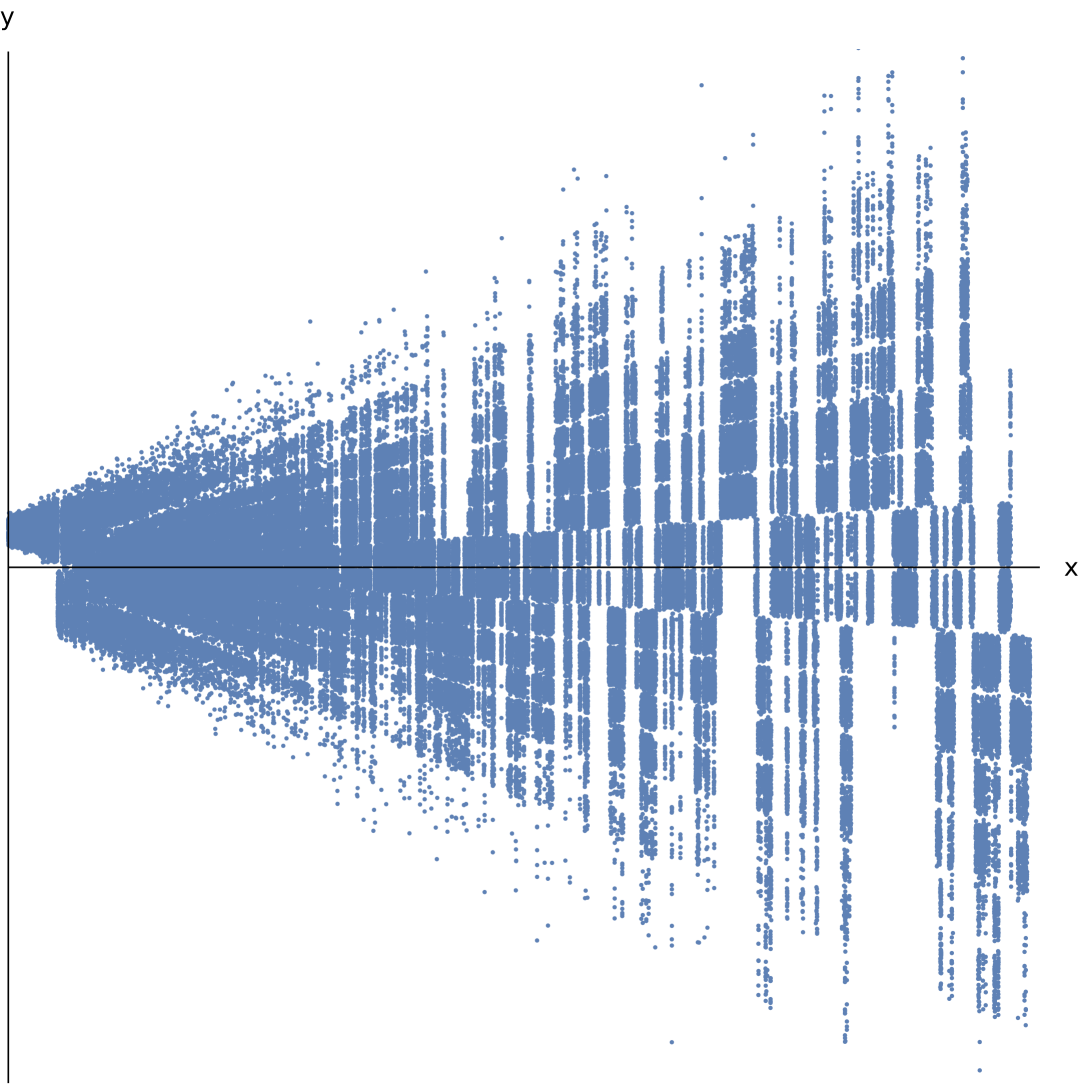}
\caption{Wall-screen distance is ca.\ \SI{1}{\milli\meter}. The vertical 
extension measures ca.\ \SI{1}{\micro\meter}.\label{1mm}}
\end{subfigure}\hfill%
\begin{subfigure}{0.45\textwidth}
\includegraphics[width=\textwidth]{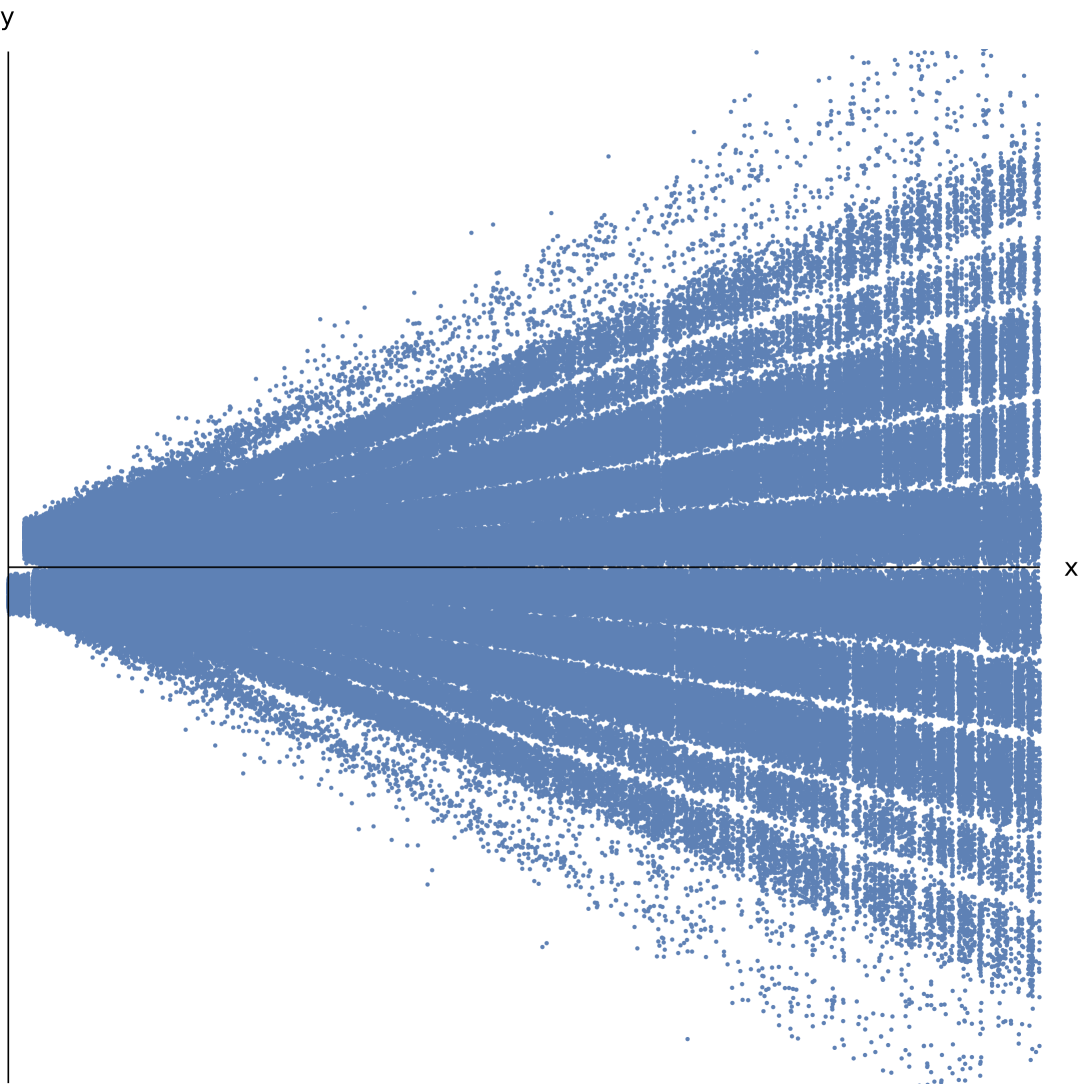}
\caption{Wall-screen distance is ca.\ \SI{1}{\meter}. The vertical extension 
measures ca.\   \SI{40}{\micro\meter}. \label{1m} }
\end{subfigure}
\caption{Trajectory of a single particle with $\vu{s} = \vu{y}$ 
traversing the given distance. Unlike in the other figures, we only plot a 
point where a tumble happens, and not the (mostly straight) parts of the 
trajectories in between, as this would completely wash out the image.}
\label{fig:long}
\end{figure}

As we scale up the physical parameters of the simulation, the zig-zag 
completely dominates the motion, as shown in Fig.~\ref{fig:long}. In Fig.~\ref{1mm}, the 
distance between the wall and the screen is $\num{3e9} \hbar / m c \approx \SI{1}{\milli\meter}$, with slits that are about \SI{15}{\nano\meter} 
wide and spaced \SI{100}{\nano\meter} apart. In Fig.~\ref{1m}, the distance 
between the wall and the screen is $\num{3e12} \hbar / m c \approx 
\SI{1}{\meter}$, with slits that are about \SI{500}{\nano\meter} wide, spaced 
\SI{3}{\micro\meter} apart.  On their way to the screen, the particles undergo 
respectively about \num{200000} and \num{5600000} tumbles. In contrast to the 
behavior on the smaller scales, the particle now explores the full width of the 
packet in the $y$-direction in both cases. While channels of probability are 
still formed by the $|\psi|^2$-distribution, trapping the particle for a 
certain amount of time, there is always a non-zero probability it will 
transition to another channel.

\section{Arrival times}\label{rem}

\begin{figure}[h!]
\begin{subfigure}{0.45\textwidth}
	\includegraphics[width=\textwidth]{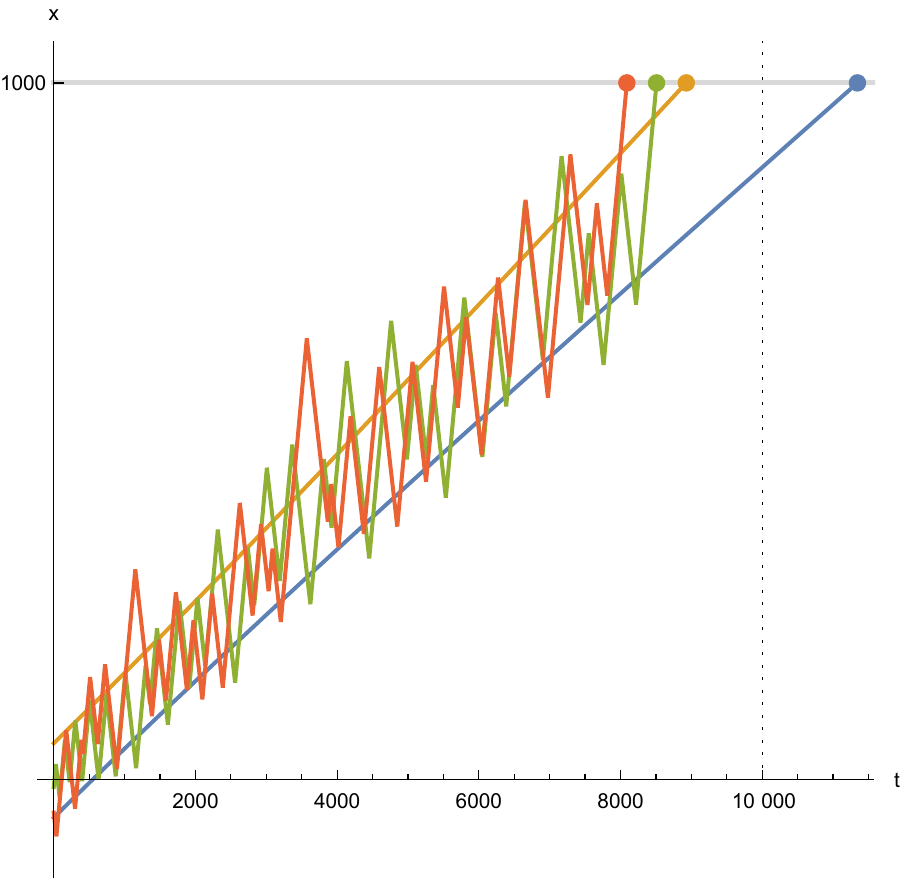}
	\caption{Two zig-zag trajectories (green, red) compared to two trajectories without curling and tumbling (blue, orange). The horizontal grey line indicates 
	the position of the screen.}
\end{subfigure}\hfill%
\begin{subfigure}{0.45\textwidth}
	\includegraphics[width=\textwidth]{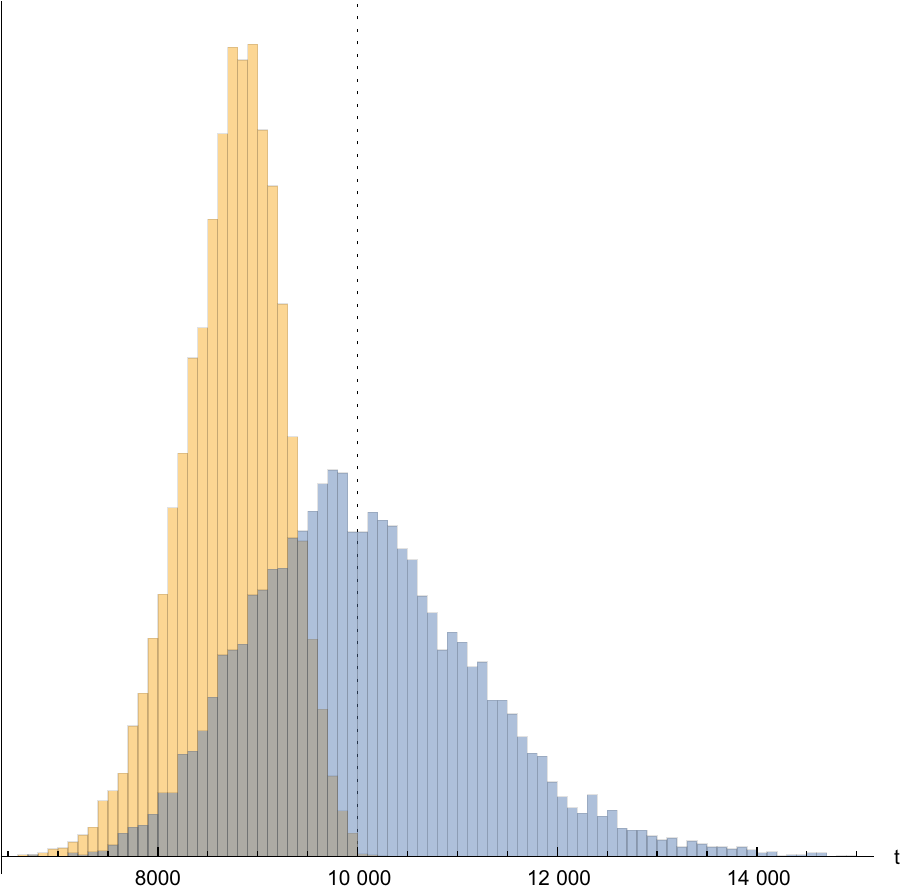}
	\caption{Comparison of the distribution of the  arrival times of zig-zag 
	trajectories (orange) with that in case of the dynamics without curling and tumbling (blue).}
\end{subfigure}
	\caption{Arrival times in the case of a single Gaussian wave function \eqref{ga} and $\vu{s} = \vu{x}$. In both plots, the dotted vertical line indicates the time at 
	which the center of the wave packet reaches the screen.}
	\label{fig:arr}
\end{figure}

An implication concerns the possibility for measuring arrival times.  
Predicting or even defining arrival times for quantum particles are subject of 
on ongoing discussion, see \cite{leavens,sidnoth,sidward} and references therein. Arrival times can however 
straightforwardly be determined in terms of the trajectories, here just by considering the time the particle first hits the screen. We consider a simple example as an initial exploration. 

While the electron moves at the speed of light between any two tumblings, the 
observed speed is less when dividing the distance between slits and screen by 
the time traveled. It is in that sense that tumbling (caused by the electron 
mass) ``decelerates'' the electron.  However, this implies the electron's 
measured velocity will have a stochastic component, or rather, the time of its 
arrival at the screen will. Such a phenomenon is always present if the spin has 
a non-zero component in the direction of movement. As an illustration, we 
consider the Gaussian wave function \eqref{ga} in the $x$-direction and take 
the spin to lie in the $x$-direction. Since the dynamics in this direction 
decouples from that in the other directions, we can ignore the $y$- and 
$z$-components of the motion. The dynamics is then given by \eqref{xcomp}.
The initial 
positions of the particles are taken according to the Born rule. In 
Fig.\ref{fig:arr} we compare the zig-zag trajectories to one without zig-zag (only first term in \eqref{vf}), and we take a look at the distribution of the 
arrival times of \num{10000} particles.  The overwhelming majority of particles 
arrive much earlier than would be the case when the curling and tumbling are omitted in 
\eqref{vf}.

It might seem puzzling that differences arise in the arrival time distributions depending on whether or not curling and tumbling are included in the dynamics. Namely, in each case, the ensemble position distribution is given by $|\psi|^2$. Therefore these theories agree on the distributions of positions of things, including e.g.\ the postions of the hands of a clock. The reason that we get different arrival times stems from the fact that we are dealing with a simplified analysis, which does not take into account the details of the measurement set-up, i.e., the details of how the particles get detected on the screen. It is striking that these effects seem to be important for time measurements, since they do not seem to matter for position measurements. In any case, including the details of the detection process in the analysis seems as yet not practically feasible, therefore it is worthwhile investigating approximative approaches like the one considered here, which ignore these details. Experiments should determine which one of the approximation methods is more successful. It would therefore be interesting to consider realistic setups as e.g.\ described in \cite{duerr19b}, to see how the arrival time distributions for the zig-zag motion compare with those when tumbling or curling are ignored. 


\section{Back to classical run-and-tumbling}\label{awa}

\begin{figure}
	\begin{subfigure}{0.45\textwidth}
		\includegraphics[width=\textwidth]{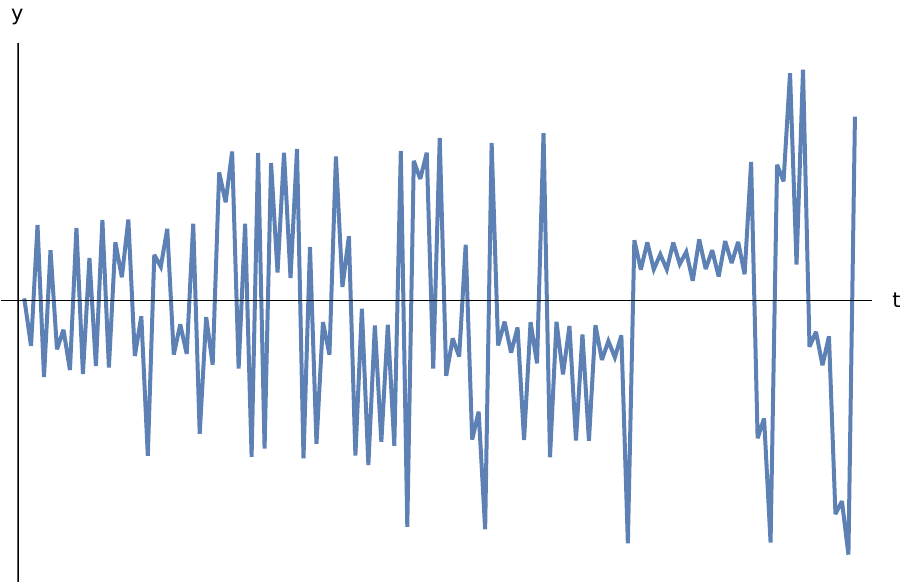}
		\caption{One-dimensional trajectory: $y$-position versus time for a 
		run-and-tumble particle moving at constant speed and tumbling according 
		to the oscillatory curve in (b).}
		\label{fig:awytraj}
	\end{subfigure}\hfill%
	\begin{subfigure}{0.45\textwidth}
		\includegraphics[width=\textwidth]{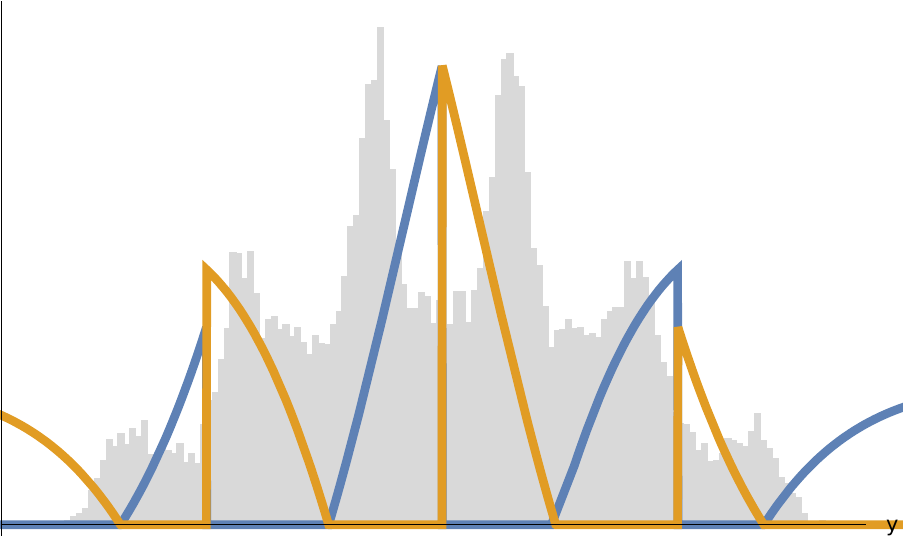}
		\caption{Distribution of final positions, and associated tumbling rates \eqref{rate}.}
		\label{fig:awyhist}
	\end{subfigure}
	\caption{Leaving quantum mechanics: the tumbling rates are given by \eqref{rate} and not 
		derived from a wave function.}
	\label{fig:awy}
\end{figure}

In the above, the trajectories realize the motion of a quantum particle, with its wave function guiding it.  The result was a stochastic dynamics caused by the flipping of chiralities.\\ 
If we leave the strict quantum prescriptions \eqref{rte} for velocity and 
tumbling fields, we remain with run-and-tumble particles as described in 
Section \ref{set}. In particular, we can use versions of tumbling rates which qualitatively 
reproduce the tumbling field Fig.~\ref{fig:ratefunctions}. For example, 
we consider a one-dimensional model where the rates are as in \eqref{eqn:rate} but
given by the positive and negative parts of
\begin{equation}
\tau(t,y) = \frac{y}{40} + 8\left(\frac{y}{12} - \left\lfloor\frac{y}{12}\right\rfloor-\frac{1}{2}\right)\, \tanh\left( \frac{t}{1000} \right)\sech\left(\frac{y}{15} \right)
\label{rate}
\end{equation}
That rate function has a linear part to keep particles from running away to 
infinity, on which we superimpose a slowly growing sawtooth, modulated by a 
hyperbolic secant envelope.  We use a run-and-tumble particle moving in the 
$y$-direction, at constant speed $\pm c$. In Fig.~\ref{fig:awytraj} we plot the 
(one-dimensional) trajectory of a run-and-tumble particle which, by locally 
increased tumbling, gets trapped (more) in certain $y$-canals. The tumbling 
rate as plotted in Fig.~\ref{fig:awyhist} follows only approximately the 
behavior of the tumbling rates
for the quantum mechanical case of Fig.~\ref{fig:ratefunctions}, but there is 
no wave function now. Therefore, the interference of fermions can be embedded in a larger class of stochastic particle dynamics 
showing these wave effects. See also \cite{loewe,burada} for similar points. 
From the point of view of soft condensed matter, it remains perhaps surprising 
that run-and-tumble particles, in a classical simulation of trajectories, may show an interference pattern.

\section{Conclusions}\label{con}
The Standard Model of particle physics suggests that electrons be viewed as run-and-tumble particles. Adopting that view in the context of the Dirac theory, analytic results for the wave function evolution can be used to determine the velocity fields and tumbling rates of the particle dynamics.  We applied that to the case of diffraction and interference, visualizing the details of the run-and-tumble process, detailing the effects of spin, while recovering the usual patterns on the screen. 
Comparing in 
Fig.~\ref{fig:doubleslittrajectories2}--Fig.~\ref{fig:doubleslittrajectories3}
with the usual trajectories for the Schr\"odinger equation (keeping only the first term on the right-hand side of \eqref{vf} and no tumbling), which lack the zig-zag dynamics, our new 
trajectories are markedly different.
We observe that 
particles in their zig-zag motion explore the probability space available to 
them, however as the interference of the wave function takes place, the 
particles tend to get trapped for longer in ``interference canals'' of high 
probability.  That can be simulated (literally now) by {\it pure} 
run-and-tumble particles (no quantum mechanics) as was shown in 
Fig.~\ref{fig:awy}.\\
 We have also started exploring arrival-time distributions, readily obtained in terms of the simulated trajectories. Interesting effects appear and it needs to be seen how these results compare to experiment. \\

\noindent {\bf Acknowledgment}:  We thank Siddhant~Das for useful discussions.  
The paper is dedicated to the memory of our dear colleague, friend and mentor 
Detlef D\"urr.

\end{document}